\documentclass[%
 reprint,
 amsmath,amssymb,
 aps,
]{revtex4-1}

\usepackage{braket}
\usepackage[mathscr]{euscript}
\usepackage[normalem]{ulem}

\usepackage{graphicx}
\usepackage{dcolumn}
\usepackage{bm}
\usepackage{color}
\usepackage{hyperref} 
\hypersetup{
    colorlinks = true,
    citecolor={blue},
    linkcolor=black,
}

\begin{document}

\title{  Entanglement negativity
  and minimal entanglement wedge cross sections
  in holographic theories}
\author{Jonah Kudler-Flam}
\email{jkudlerflam@uchicago.edu}
\author{Shinsei Ryu}%
 \email{ryuu@uchicago.edu}
\affiliation{%
 James Franck Institute and Kadanoff Center for Theoretical Physics\\ University of Chicago, Illinois 60637, USA.
}%

\date{\today}

\begin{abstract}
  We calculate logarithmic negativity,
  a quantum entanglement measure for mixed quantum states, 
  in quantum error-correcting codes and find it to equal the minimal cross
  sectional area of the entanglement wedge
  in holographic codes with a quantum correction term equal to the logarithmic
  negativity between the bulk
  degrees of freedom on either side of the entanglement wedge cross section.
  This leads us to conjecture a holographic dual for logarithmic negativity
  that is related to the area of a cosmic brane with tension in the entanglement
  wedge plus a quantum correction term. This is closely related to (though distinct from) the holographic proposal for entanglement of purification.
  We check this relation for various configurations of subregions in ${\it AdS}_3/{\it CFT}_2$. These are disjoint intervals at zero temperature, as well as a single interval and adjacent intervals at finite temperature. We also find this prescription to effectively characterize the thermofield double state. We discuss how a deformation of a spherical entangling region complicates calculations and speculate how to generalize to a covariant description.
\end{abstract}

\maketitle

\tableofcontents

\section{Introduction}

Holographic duality (the AdS/CFT correspondence)
\cite{1999IJTP...38.1113M,1998AdTMP...2..253W, 1998PhLB..428..105G}
has made a dramatic impact on how we understand
theories of quantum gravity and strongly coupled conformal field theories.
One of the recent explorations in this context 
is a connection to quantum information theory,
aiming to uncover 
the mechanism of holographic duality and quantum gravity
\cite{2009JPhA...42X4008N,2017LNP...931.....R,2018arXiv180110352N, 2014arXiv1409.1231H}.
In particular, it has been proposed that the duality can be interpreted as a quantum
error-correcting code \cite{2017CMaPh.354..865H,2015JHEP...04..163A,2015JHEP...06..149P}.
This surprising connection
has been able to shed light on mysterious parts of
holographic duality.
For example, 
it helps to explain the holographic formula of entanglement entropy, 
which equates
the von Neumann entropy
of the boundary conformal field theory (CFT) to the geometry of the bulk AdS \cite{2006PhRvL..96r1602R,2006JHEP...08..045R}:
\begin{align}
  S(\rho_A) = \frac{ {\it Area}(\mathscr{L}_A)}{4 G_N}
  + S_{{\it bulk}}.
\label{RT}
\end{align}
Here, $S(\rho_A)$ is the von Neumann entropy
of the subregion $A$ in the boundary CFT,
and $\mathscr{L}_A$ is the extremal surface in the bulk homologous to $A$;
$G_N$ is the bulk Newton constant.
$S_{{\it bulk}}$ is the bulk entanglement entropy of the corresponding
entanglement wedge,
the quantum correction term \cite{2013JHEP...11..074F}.
This formula was shown to hold in the case of ``holographic states'' made of
perfect tensors \cite{2015JHEP...06..149P} and in random tensor networks \cite{2016JHEP...11..009H}.
It was later proven more generally for quantum error-correcting codes
\cite{2017CMaPh.354..865H}.

For mixed quantum states, the von Neumann entropy
is not a proper measure for the quantum correlation;
it captures classical (thermal) correlations as well as purely
quantum ones.
The (logarithmic) entanglement negativity is a measure of quantum entanglement,
which can be applied to mixed states.
In the quantum field theory context,
for example, 
it has been computed and discussed for
(1+1)d CFTs 
and
(2+1)d topological quantum field theories
\cite{
2012PhRvL.109m0502C,
2013JSMTE..02..008C,
2013JSMTE..05..002C,
2013JSMTE..05..013A,
PhysRevB.90.064401,
2016PhRvB..94c5152R,
2016PhRvB..94s5121R,
2016JPhA...49l5401B,
2014NJPh...16l3020E,
2014JSMTE..12..017C,
2015NuPhB.898...78H,
PhysRevB.92.075109,
PhysRevA.88.042319,
PhysRevA.88.042318,
2016JHEP...09..012W,
2016PhRvB..93x5140W}.

In this paper, we will make a comparison,
in the holographic context, 
between 
the entanglement negativity
and
the minimal cross sectional area of
entanglement wedges.
The entanglement wedge has been proposed as a natural bulk
region corresponding to a given boundary region \cite{2014JHEP...12..162H,2014arXiv1412.8465J,2016JHEP...06..004J}
and has proven to be an important concept, distinct from the causal wedge,
when discussing bulk reconstruction \cite{2015JHEP...06..149P, 2015JHEP...04..163A, 2014arXiv1412.8465J, 2016JHEP...06..004J,2012JHEP...06..114H,2014JHEP...12..162H}.
In addition, it was recently proposed that minimal entanglement wedge cross
sections are a measure of the entanglement of purification (EoP)
\cite{2018NatPh..14..573U,2018JHEP...04..132B, 2018JHEP...01..098N, 2018PTEP.2018f3B03H,2018JHEP...03..006B,2018arXiv180405855E, 2018arXiv180500476B,2018JHEP...10..152U, 2018PhRvD..98b6010N}.
We will discuss the distinctions between the bulk objects proposed as duals to the logarithmic negativity and EoP.
It is also worth mentioning that there is a proposal for holographic negativity that relates certain combinations of bulk geodesics to the negativity in the boundary CFT \cite{2017arXiv170708293J, 2016arXiv160201147C, 2018EPJP..133..300J, 2016arXiv160906609C, 2019NuPhB.94514683J}. We find inconsistencies between bulk and boundary computations for this proposal.

First, we consider the logarithmic negativity in generic quantum error-correcting codes. With this formalism, we study negativity and the entanglement wedge cross section in holographic quantum error-correcting codes -- they are toy models of holographic duality. There, as we will see, the logarithmic negativity is equivalent to the cross sectional area of entanglement wedges with a bulk quantum correction term.  
We explicitly show this for setups where we bipartition the (boundary) system at finite temperature.

With motivations from quantum error-correcting codes, we next conjecture a general bulk object that computes the logarithmic negativity. While for general entangling surface geometries, this is difficult to compute due to the backreaction of the cosmic branes that we will introduce, the calculation is greatly simplified for ball shaped subregions. In these symmetric set-ups, the backreaction is accounted for by an overall constant to the area of the entanglement wedge cross section such that the logarithmic negativity, $\mathcal{E}$, in holographic CFTs
is given by
\begin{align}
  \label{holgraphic negativity formula AdS3/CFT2}
  \mathscr{E} = \mathcal{X}_d \frac{E_W}{4G_N} + \mathcal{E}_{bulk},
\end{align}
where $E_W$
is the minimal cross sectional area of the entanglement wedge associated with the boundary region of interest and $\mathcal{X}_d$ is a constant which depends on the dimension of the spacetime. $\mathcal{E}_{bulk}$ is the quantum correction term corresponding to the logarithmic negativity between the bulk fields on either side of the cross section. In ${\it AdS}_3/{\it CFT}_2$ (where $\mathcal{X}_d = 3/2$), we find that the entanglement wedge cross section formula reproduces many known properties of the logarithmic negativity in (1+1)d (holographic) CFTs. 

For the rest of the introduction,
we briefly review the definitions of
the key concepts in this paper;
the (logarithmic) entanglement negativity,
the entanglement wedge, and
holographic error-correcting codes.

\subsection{Entanglement negativity}

For bipartite pure states,
the von Neumann entropy of the reduced density matrix effectively encapsulates
the quantum correlations between subsystems.
However, when working with mixed states,
the von Neumann entropy is not a proper entanglement measure;
for example, the von Neumann entropy additionally
counts the classical correlations.
In particular, in thermal systems without quantum correlations,
this will just be the regular thermal entropy.

The (logarithmic) negativity was proposed as a computable measure of quantum
entanglement for mixed states \cite{1996PhRvL..77.1413P, 1996PhLA..223....1H, 1999JMOp...46..145E, 2002PhRvA..65c2314V, 2005PhRvL..95i0503P}.
The negativity is a measure derived from
the positive partial transpose (PPT) criterion for the separability
of mixed states \cite{1996PhRvL..77.1413P},
and is defined/computed
by taking the trace norm of the partial transpose of the density matrix:
For the Hilbert space $\mathcal{H}_A \otimes \mathcal{H}_B$,
the partial transpose of the density matrix $\rho$
is defined in terms of its matrix elements as 
\begin{equation}
\bra{i_A,j_B} \rho^{T_A} \ket{k_A, l_B} = \bra{k_A, j_B}\rho\ket{i_A, l_B},
\end{equation}
where $\{|i_{A/B}\rangle\}$
represent the orthonormal bases for $\mathcal{H}_{A/B}$.
The entanglement negativity and logarithmic negativity are defined as 
\begin{align}
  &
  \mathscr{N}(\rho) := \frac{1}{2}
  \left(
    \left|\rho^{T_A} \right|_1 -1 
  \right),
   \nonumber \\
  &
  \mathscr{E}(\rho) := \log\left|\rho^{T_A} \right|_1,
\end{align}
where
$\left| A \right|_1 := \mathrm{Tr}\, \sqrt{A A^{\dag}}$.
In this paper, we will be mainly concerned with
the logarithmic negativity
(and hence
by entanglement negativity,
we refer to $\mathscr{E}$
unless stated otherwise).

\subsection{Entanglement wedge}
The entanglement wedge is the bulk region corresponding to the
reduced density matrix on the boundary.
In this paper,
we are only concerned with entanglement of the CFT on the boundary on a fixed time slice, corresponding to a Cauchy slice of $AdS$ in the bulk.
Relevant generalizations of entanglement entropy to time-dependent situations have been studied in Refs.~\cite{2007JHEP...07..062H, 2014CQGra..31v5007W}.
Given a Cauchy slice, $\Xi$,
and a subset of the conformal boundary, $A \subset \partial \Xi$, the
relevant surface, $\gamma_A$, is the codimension-2 extremal surface homologous
to $A$, $\partial A = \partial \gamma_A$.
The corresponding entanglement wedge of $A$ is the codimension-1 surface in
$\Xi$ whose boundary is $\gamma_A \cup A$.

\begin{figure}
    \includegraphics[width=.35\linewidth]{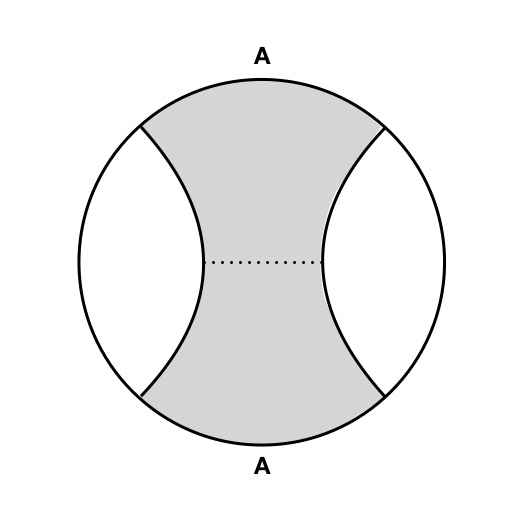}
    \includegraphics[width=.35\linewidth]{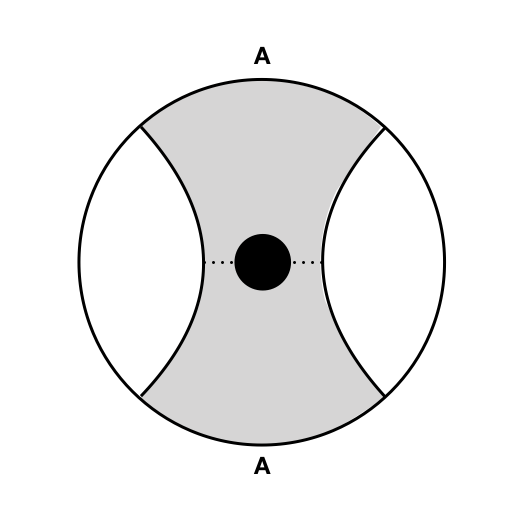}
\caption{The gray bulk region is the entanglement wedge of boundary subregion
  $A$. The dotted line represents the minimal entanglement wedge cross section.
  The figure on the right displays a black hole. The cross section then becomes
  disconnected, containing pieces on either side of the black hole but not
  including any of the horizon.
}
\label{ent_wedge}
\end{figure}

We are interested in the minimal cross sectional area of the entanglement wedge. This picture is intuitive when the bulk does not contain horizons. However, when there is a black hole in the bulk, the entanglement wedge cross section can become disconnected (Fig.~\ref{ent_wedge}).
\subsection{Holographic codes}

\begin{figure}
    \includegraphics[width = 0.4 \columnwidth]{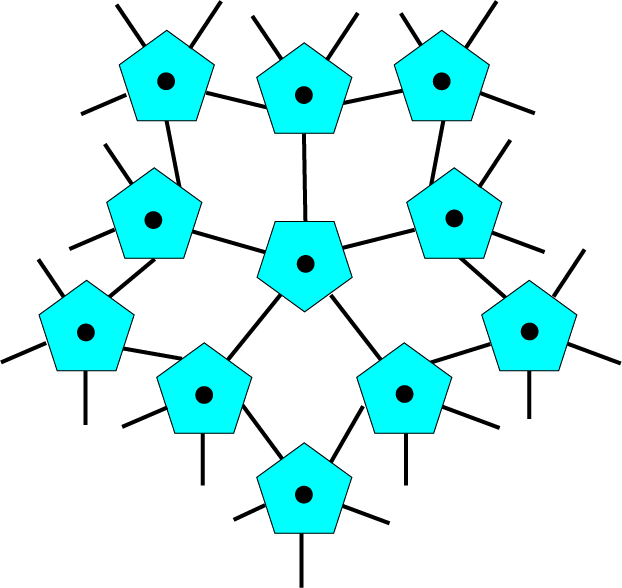}
    \caption{The holographic pentagon code introduced in
      Ref.~\cite{2015JHEP...06..149P}. Each perfect tensor,
      represented by a pentagon, has six indices, with one free bulk
      index (represented by dots). 
      }
    \label{pentagon_code}
\end{figure}

A series of concrete and exactly solvable toy models of holography,
holographic codes, 
were proposed in Ref.~\cite{2015JHEP...06..149P}.
Leveraging the fact that the AdS/CFT correspondence shares central properties
with quantum error-correcting codes, the authors studied a tensor network
description of a quantum-error-correcting code living on a given two-dimensional
time slice.
This code acts as an isometric map
from the bulk Hilbert space to the boundary Hilbert space. This ``holographic code" is composed of perfect tensors which are
tensors such that any partition of indices into  $\mathcal{H}_A$ and $\mathcal{H}_B$ induces an isometry $T$ from $\mathcal{H}_A$ to $\mathcal{H}_B$, given that $\vert \mathcal{H}_A \vert \leq \vert \mathcal{H}_B \vert$.
An example of a holographic code model,
the holographic pentagon code,
is depicted in Fig.~\ref{pentagon_code}.
For a complete discussion of such codes, see Ref.~\cite{2015JHEP...06..149P}.

\section{Entanglement wedge and negativity in quantum error-correcting codes}
In this section, we will calculate the logarithmic negativity in generic quantum error correcting codes. We will later use this technology to gain geometrical insight into logarithmic negativity in holographic codes.
Following the structure of Ref.~\cite{2017CMaPh.354..865H},
we warm up by starting with simple erasure correcting code models for holography and then continue to more general error-correcting codes.

\subsection{Conventional QEC}
\label{Conventional QEC}

We work with a total Hilbert space
$\mathcal{H}$ endowed with the tensor product structure
$\mathcal{H} = \mathcal{H}_A \otimes \mathcal{H}_{\bar{A}}$;
$\mathcal{H}_A = (\mathcal{H}_{A_1} \otimes \mathcal{H}_{A_2}) \oplus
\mathcal{H}_{A_3}$.
The logical state is then encoded in the state subspace such that there exists a unitary $U_A$ such that
\begin{align}
  \ket{\tilde{i}} &= U^{\ }_A ( \ket{i}_{A_1} \otimes \ket{\chi}_{A_2,\bar{A}}),
                    \quad
\ket{\chi}_{A_2,\bar{A}} \in \mathcal{H}_{A_2, \bar{A}}.
\end{align}
This implies a code that corrects for the erasure of $\bar{A}$. 
Here, the state $|\chi\rangle_{A_2, \bar{A}}$ is our entanglement resource for quantum error correction.

The 3-qutrit code is the simplest example of conventional quantum error-correction that displays holographic properties.
It consists of three physical qutrits,
each with states $\ket{0}$, $\ket{1}$, and $\ket{2}$
that encode a single logical qutrit $\ket{\tilde{i}}$ as follows:
\begin{align}
  \ket{\tilde{0}} &= \frac{1}{\sqrt{3}} ( \ket{000} + \ket{111} + \ket{222}),
                     \nonumber \\
  \ket{\tilde{1}} &= \frac{1}{\sqrt{3}} ( \ket{012} + \ket{120} + \ket{201}),
                     \nonumber \\
\ket{\tilde{2}} &= \frac{1}{\sqrt{3}} ( \ket{021} + \ket{102} + \ket{210}).
                  \label{3 qutrit code}
\end{align}
This code can correct for the erasure of any single physical qutrit because
there exists a unitary operator, $U_{A_1 A_2}$,
such that
\begin{align}
  U^{\dagger}_{A_1 A_2} \ket{\tilde{i}} &= \ket{i}_{A_1} \ket{\chi}_{A_2 \bar{A}}, \quad
\ket{\chi} \equiv \frac{1}{\sqrt{3}}(\ket{00} +\ket{11} + \ket{22}),
\end{align}
where $A_1$, $A_2$, and $\bar{A}$ correspond to the three qutrits. See Ref.~\cite{2015JHEP...04..163A} for the explicit $U_{A}$. Because of the symmetry of the
code, there also exist analogous unitary operators $U_{A_2 \bar{A}}$ and $U_{A_1 \bar{A}}$.
In Refs.~\cite{2015JHEP...04..163A,2017CMaPh.354..865H}, this simple code is shown to contain analogs of
many important aspects of holography: black holes,
effective field theory,
radial commutativity,
subregion duality, and
the holographic formula of entanglement entropy
(the RT formula).

We will only replicate the argument for
the RT formula here because the other properties do not directly apply to the calculation and interpretation of the negativity of this code.  
Consider an arbitrary mixed state of a conventional quantum error-correcting code
\begin{align}
  \tilde{\rho} &= U^{\ }_{A_1 A_2} ( \rho^{\ }_{A_1} \otimes \ket{\chi}\bra{\chi}_{A_2, \bar{A}})U_{A_1 A_2}^{\dagger},
\end{align}
where $\rho_{A_1}$ is an arbitrary mixed input state. Defining $\chi_{A_2} \equiv \mathrm{Tr}_{\bar{A}}\, \ket{\chi}\bra{\chi}_{A_2,
  \bar{A}}$,
the von Neumann entropies for the reduced density matrices
$\tilde{\rho}_A = \mathrm{Tr}_{\bar{A}}\, \tilde{\rho}$ 
and 
$\tilde{\rho}_{\bar{A}} = \mathrm{Tr}_{A}\, \tilde{\rho}$ 
are 
\begin{align}
  S(\tilde{\rho}_A) &= S(\chi_{A_2}) + S(\tilde{\rho}),
                      \quad
  S(\tilde{\rho}_{\bar{A}}) =  S(\chi_{A_2}).
  \label{conventional_EE}
\end{align}
By identifying $S(\chi_{A_2}) I_{{\it code}}$
as the ``area operator," $\mathscr{L}$,
\begin{align}
  \langle \mathscr{L} \rangle=
  S(\chi_{A_2})
  =
  - \sum_a p_a \log{p_a},
\end{align}
an RT-like formula for error-correcting codes is obtained. $\mathcal{L}$ can be thought of as an area because it contributes equally to $A$ and $\bar{A}$. Furthermore, if one works with tensor networks, $\mathcal{L}$ originates from $\ket{\chi}$ which make up the Hilbert space of the contracted legs of the network. Though this initial
formulation of error-correcting codes displays certain aspects of holography, it
is not entirely satisfactory. This is partially due to the entanglement entropy not being
symmetric. Only for system $A$ is there a bulk entropy term. In the next
section, we expand to more general error-correcting codes so that both entropies
contain bulk entropy terms, as we expect they should. Another motivation for
this generalization is that we will be able to apply our results to the
holographic codes introduced in Ref.~\cite{2015JHEP...06..149P}.

\begin{figure}[ht]
\includegraphics[width = 0.95 \columnwidth]{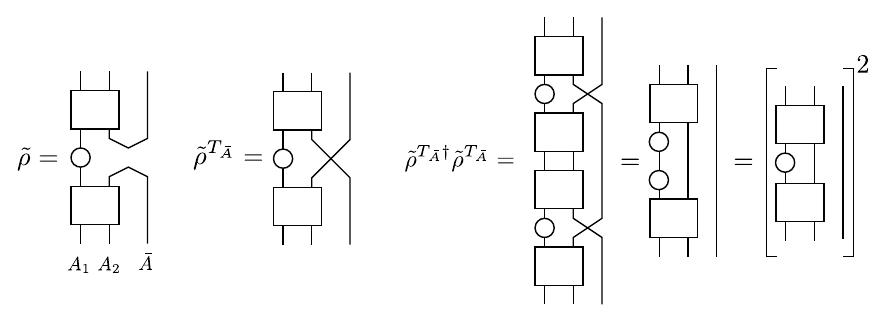}
\caption{
Graphical representations of Eq.~\eqref{part trans}.
Here, squares represent $U_A$ or $U_A^{\dag}$
and circles represent $\rho_{A_1}$
}
\label{qec-neg}
\end{figure}

Before discussing more generic holographic code models,
let us consider the negativity of the conventional QEC model.
In order to take the partial transpose with respect to $A$ or $\bar{A}$, we need to perform a Schmidt decomposition
of $|\chi\rangle$:
\begin{align}
\ket{\chi} &= \sum_a \sqrt{p_a} \ket{a}_{A_2} \otimes \ket{a}_{\bar{A}}.
\end{align}
$\ket{\chi}$ is maximally entangled when
\begin{align}
    p_a =  \frac{1}{\vert \tilde{A} \vert},
\quad
\vert \tilde{A} \vert = \mbox{min}(\vert A_2 \vert, \vert \bar{A} \vert).
\end{align}
Taking the partial transpose with respect to $\bar{A}$
\footnote{We would end up with the same result for negativity if we took the partial
transpose with respect to $A$ because the negativity is symmetric about
bipartite states.},
\begin{align}
  \label{part trans}
  \tilde{\rho}^{T_{\bar{A}}}
  &= \sum_{a,b} \sqrt{p_a p_b}\, U^{\ }_{A_1 A_2} ( \rho_{A_1} \otimes \ket{a}\bra{b}_{A_2}  \otimes \ket{b}\bra{a}_{\bar{A}})U_{A_1 A_2}^{\dagger},
    \nonumber \\
  \tilde{\rho}^{{T_{\bar{A}}} \dagger} \tilde{\rho}^{T_{\bar{A}}}
&= \sum_{a,b} p_a p_b\, U^{\ }_{A_1 A_2} (\rho_{A_1}^2 \otimes \ket{a}\bra{a}_{A_2} \otimes \ket{b}\bra{b}_{\bar{A}} )U_{A_1 A_2}^{\dagger},
     \nonumber \\
    \left| \tilde{\rho}^{T_{\bar{A}}} \right|_1
&= \Big(\sum_a \sqrt{p_a}\Big)^2. 
\end{align}
(See Fig.~\ref{qec-neg} for graphical representations of these objects,
when $|\chi\rangle$ is maximally entangled.)
We obtain the entanglement negativity $\mathscr{N}$ and logarithmic negativity $\mathscr{E}$
\begin{align}
\mathscr{N}(\tilde{\rho}) &= \frac{\Big(\sum_a \sqrt{p_a}\Big)^2 - 1}{2},
  \\
\mathscr{E}(\tilde{\rho}) &= \log{\Big(\sum_a \sqrt{p_a}\Big)^2} = S_{1/2}(\chi_{A_2}),
\end{align}
where $S_{1/2}$ is the R\'enyi entropy with R\'enyi index $1/2$. So the negativity is equal to the expectation value of the area operator
$\langle  \mathscr{L} \rangle$
when $\chi_{A_2}$ is maximally mixed:
\begin{align}
  \mathscr{E}(\tilde{\rho})
  &= \langle \mathscr{L} \rangle = \log( \vert \tilde{A}\vert).
\label{init_neg}
\end{align}
For tensor networks, because the spectrum of the entanglement Hamiltonian
is flat,  $\chi_{A_2}$ is maximally mixed and we find no difference from the von Neumann entropy. However, when we move to AdS/CFT, the spectrum is not flat and this term accounts for the tension of the cosmic brane. These codes are also not entirely satisfactory because there is no quantum correction to the logarithmic negativity. 

\begin{figure}[ht]
 
\includegraphics[width = 0.3 \columnwidth]{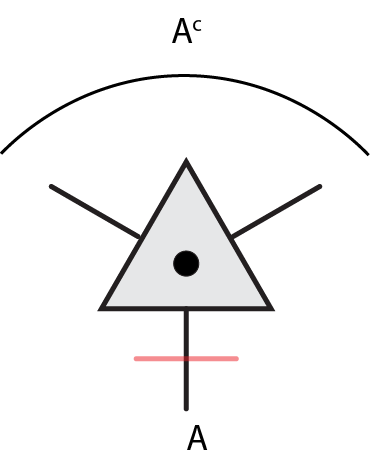}
\caption{The tensor network representation of the 3-qutrit code.
  There is only one tensor in this network. It maps the single bulk logical qutrit (central black dot) to the three physical qutrits. The red line represents the minimal geodesic separating boundary region $A$ and its complement, $A^c$. }

\label{qutrit_fig}
\end{figure}

Because $\ket{\chi}$ for the 3-qutrit code is maximally entangled, we can apply \eqref{init_neg}. When bi-partitioning the boundary, the bulk minimal geodesic cuts only a single leg (Fig.~\ref{qutrit_fig}) of dimension 3, leading to a negativity of $\log{(3)}$.

\subsection{Subsystem QEC with complementary recovery}
\label{subsystem_section}
\begin{figure}
    \centering
    \includegraphics[height = 3cm]{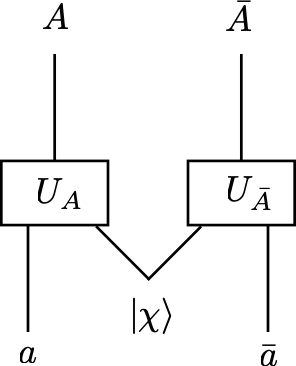}
    \caption{In subsystem quantum error correction with complementary recovery, ``bulk" degrees of freedom in the code subspace ($\mathcal{H}_a \otimes \mathcal{H}_{\bar{a}}$) are encoded in the ``boundary" Hilbert space ($\mathcal{H}_A \otimes \mathcal{H}_{\bar{A}}$) using the auxiliary state $\ket{\chi}$ as an entanglement resource.}
    \label{subsystem_diagram}
\end{figure}

Subsystem quantum error-correction is a generalization to conventional quantum error-correction. This generalization is crucial to our analysis because the holographic codes that we will employ belong to this family of error-correcting codes. There is a further generalization that is referred to as operator-algebra quantum error-correcting codes, though we leave this analysis to future work \footnote{We note that an earlier version of this paper had an error in the generalization to operator algebra error-correcting codes.}.
Again, we will make the Hilbert space $\mathcal{H}$ factorize into $\mathcal{H}_A \otimes \mathcal{H}_{\bar{A}}$, while the code subspace factorizes as $\mathcal{H}_{code} = \mathcal{H}_a \otimes \mathcal{H}_{\bar{a}}$. This code subspace is created such that the state can be recovered either on $A$ or $\bar{A}$. This construction allows the RT formula to be symmetric. The codespace is spanned by
\begin{equation}
  \ket{\tilde{ij}} = U^{\ }_A (\ket{i}_{A_1}\ket{\chi_j}_{A_2, \bar{A}})
  = U^{\ }_{\bar{A}} (\ket{j}_{\bar{A}_1}\ket{\chi_i}_{\bar{A}_2, A}).
\end{equation}
We can simplify the code subspace to 
\begin{equation}
\label{subsystem}
\ket{\tilde{ij}} = U^{\ }_A U^{\ }_{\bar{A}} (\ket{i}_{A_1} \ket{j}_{\bar{A}_1}\ket{\chi}_{A_2 \bar{A}_2})
\end{equation}
because
\begin{align}
  U_{\bar{A}}^{\dagger} \ket{\chi_j}_{A_2, \bar{A}}
  &= \ket{j}_{\bar{A}_1}\ket{\chi}_{A_2 \bar{A}_2},
    \nonumber \\
    U_{A}^{\dagger} \ket{\chi_i}_{\bar{A}_2, A}
    &= \ket{i}_{A_1}\ket{\chi}_{A_2 \bar{A}_2}.
\end{align}
Therefore, a density matrix can be encoded as
\begin{equation}
  \tilde{\rho} = U_A U_{\bar{A}}\left(
    \rho_{A_1 \bar{A}_1}  \otimes \ket{\chi}\bra{\chi}_{A_2 \bar{A}_2}
    \right) U_{\bar{A}}^{\dagger}U_A^{\dagger}.  
\end{equation}
See Fig.~\ref{subsystem_diagram} for a graphical representation.
By defining $\chi_{A_2} \equiv \mathrm{Tr}_{\bar{A}_2}\, \ket{\chi}\bra{\chi}$
and $\chi_{\bar{A}_2} \equiv \mathrm{Tr}_{A_2}\, \ket{\chi}\bra{\chi}$,
we obtain 
\begin{align}
  \tilde{\rho}_A &= U^{\ }_A (\rho_{A_1} \otimes \chi_{A_2})U^{\dagger}_A,
        \nonumber \\
\tilde{\rho}_{\bar{A}} &= U^{\ }_{\bar{A}} (\rho_{\bar{A}_1} \otimes \chi_{\bar{A}_2})U^{\dagger}_{\bar{A}}.
\end{align}
The associated area operators are
\begin{align}
  \mathscr{L}_A \equiv S(\chi_{A_2})I_a,
  \quad
\mathscr{L}_{\bar{A}} \equiv S(\chi_{A_2})I_{\bar{a}},
\end{align}
so the new RT formulas are symmetric:
\begin{align}
S(\tilde{\rho}_A) = \langle \mathscr{L}_A \rangle + S(\tilde{\rho}_a),
  \quad
S(\tilde{\rho}_{\bar{A}}) = \langle \mathscr{L}_{\bar{A}} \rangle + S(\tilde{\rho}_{\bar{a}}).
\end{align}

The calculation of negativity in subsystem quantum error-correction with complementary recovery is quite similar to that of conventional QEC. 
We Schmidt decompose $\ket{\chi}_{A_2 \bar{A}_2}$ as
\begin{align}
&\ket{\chi}_{A_2 \bar{A}_2} = \sum_a \sqrt{p_a} \ket{a}_{A_2} \ket{a}_{\bar{A}_2},
\end{align}
so that the density matrix is
\begin{align}
    \tilde{\rho} = U_A U_{\bar{A}}\Bigg(
    \rho_{A_1 \bar{A}_1}  &\otimes \sum_{ab} \sqrt{p_a p_b} \ket{a}\bra{b}_{A_2} \\ \nonumber &\otimes \ket{a}\bra{b}_{\bar{A}_2}
    \Bigg) U_{\bar{A}}^{\dagger}U_A^{\dagger}.  
\end{align}
We now take the partial transpose with respect to $\bar{A}$
\begin{align}
    \tilde{\rho}^{T_{\bar{A}}} = U^{\ }_A U_{\bar{A}}^T\Bigg(
    \rho_{A_1 \bar{A}_1}^{T_{\bar{A_1}}}  &\otimes \sum_{ab} \sqrt{p_a p_b} \ket{a}\bra{b}_{A_2}  \\ \nonumber &\otimes \ket{b}\bra{a}_{\bar{A}_2}
    \Bigg) U_{\bar{A}}^{\dagger T} U_A^{\dagger}.  
\end{align}
Taking the trace norm, we find
\begin{align}
    \left| \tilde{\rho}^{T_{\bar{A}}}  \right|_1 = \left( \sum_a \sqrt{p_a}\right)^2 \left|\rho_{A_1 \bar{A}_1}^{T_{\bar{A_1}}}  \right|_1.
\end{align}
Therefore,
\begin{align}
  \mathcal{E}(\tilde{\rho}) &= S_{1/2}(\chi_{A_2}) + \mathcal{E}(\rho_{A_1, \bar{A}_1})
                              \label{main}\\ \nonumber
    &=  \langle \mathcal{L} \rangle  + \mathcal{E}(\rho_{A_1, \bar{A}_1}),
\end{align}
because $\chi_{A_2}$ is maximally mixed. Again, we have found the negativity to have a contribution from the area operator. However, this time there is an additional quantum correction term equal to the negativity of the input state.
We have thus found a quantum corrected holographic logarithmic negativity formula. The quantum correction term is negligible when the bulk correction to the holographic von Neumann entropy (\ref{RT}) is negligible. We again note that the appearance of $S_{1/2}$ will imply nontrivial backreaction when we move to AdS/CFT.

\subsection{Entanglement negativity in
  holographic perfect tensor network codes}
\label{Entanglement negativity in holographic perfect tensor network codes}

So far, we have been working abstractly in the language of erasure-error-correcting codes.
In order to obtain ``geometric'' insights of the entanglement structure of our
quantum states,  
we now apply the results to the holographic perfect tensor network codes introduced in Ref.~\cite{2015JHEP...06..149P}.

Holographic perfect tensor network codes are subsystem quantum-error-correcting
codes made out of perfect tensors.
They act as maps from the bulk Hilbert space of logical indices to the boundary
Hilbert space of physical indices.
The authors of Ref.~\cite{2015JHEP...06..149P} were able to analyze these codes from the
perspective of a discrete RT formula by implementing the ``greedy algorithm"
which gives a corresponding ``greedy geodesic."
The greedy geodesic is initialized at a boundary subspace $A$.
The greedy algorithm is implemented by removing tensors in the bulk one by one
if more indices of that tensor lie outside of the greedy geodesic than inside.
On a graph of negative curvature,
this process will stop at some equilibrium position within the bulk, defining the greedy geodesic $\gamma_A$ for boundary subspace $A$. The graph version of the entanglement wedge is then the union of the tensors that are bounded by $A$ and $\gamma_A$.

In the following, we will analyze the logarithmic negativity of
the holographic perfect tensor network codes.
As the usefulness of negativity arises when working with mixed states,
once again,
we are mainly interested in the following two setups:
(i) we start with a mixed state in the total (boundary) Hilbert space, and
then bipartition the boundary Hilbert space and discuss the entanglement
negativity of the bipartition, 
and
(ii) we start with a pure state but trace out a sub-Hilbert space to
obtain a mixed state for the compliment.
We then bipartition the remaining Hilbert space and discuss the entanglement
negativity. The first setup has a straightforward and illuminating answer, while the second setup is more complicated and less clear.

\subsubsection{Bipartite entanglement at finite temperature}

\begin{figure}

    \includegraphics[width = 4cm]{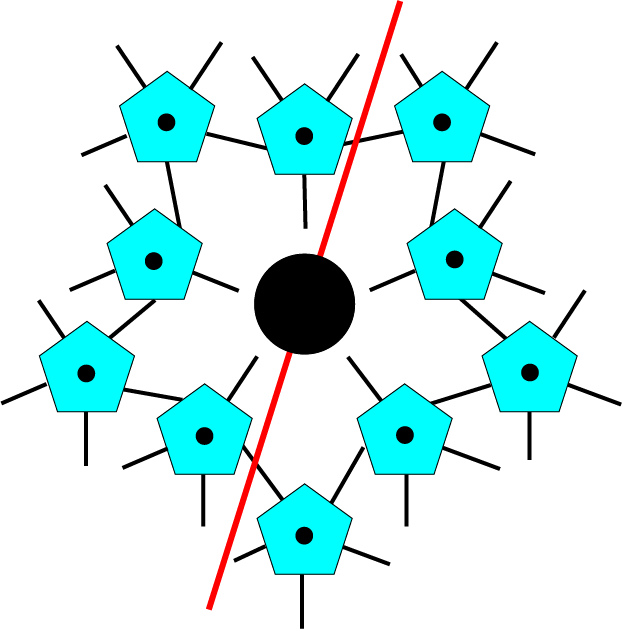}
    \caption{A black hole in a holographic code is implemented by removing the central tensor of the network. The minimal geodesic (red) homologous to $A$ does not pick up any contributions from the black hole horizon and represents the entanglement wedge cross section.}
    \label{happy_bh}
\end{figure}

For the first setup,
we put our boundary theory at finite temperature by introducing a black hole in
the center of the bulk (see Fig.~\ref{happy_bh}).
Following Refs.~\cite{2015JHEP...06..149P, 2017CMaPh.354..865H}, we implement the black hole by removing the
central tensor.
The new central legs are bulk indices that model the black hole entropy.
The resulting tensor network is a subsystem quantum-error-correcting code.
Therefore, we are able to apply the result from \eqref{main} to calculate the
negativity of the bipartition.
For the case of the entanglement entropy,
the minimal cut homologous to $A$ ``goes through" the black hole,
and hence the
entanglement entropy receives two types of contributions;
the ``quantum'' part contributed from the part of the cut which does not
``touch'' the horizon,
and the ``thermal'' part coming from the horizon.
For the case of the entanglement negativity,
\eqref{main} suggests that we simply remove the thermal contribution, and
consequently, it does not pick up the volume law contribution from the horizon.
This tensor network picture of finite temperature
holographic codes
resembles the minimal entanglement wedge area of BTZ black holes (Fig.~\ref{ent_wedge}).

\subsubsection{Tripartite entanglement}

\begin{figure}
       \includegraphics[width = 0.8 \columnwidth]{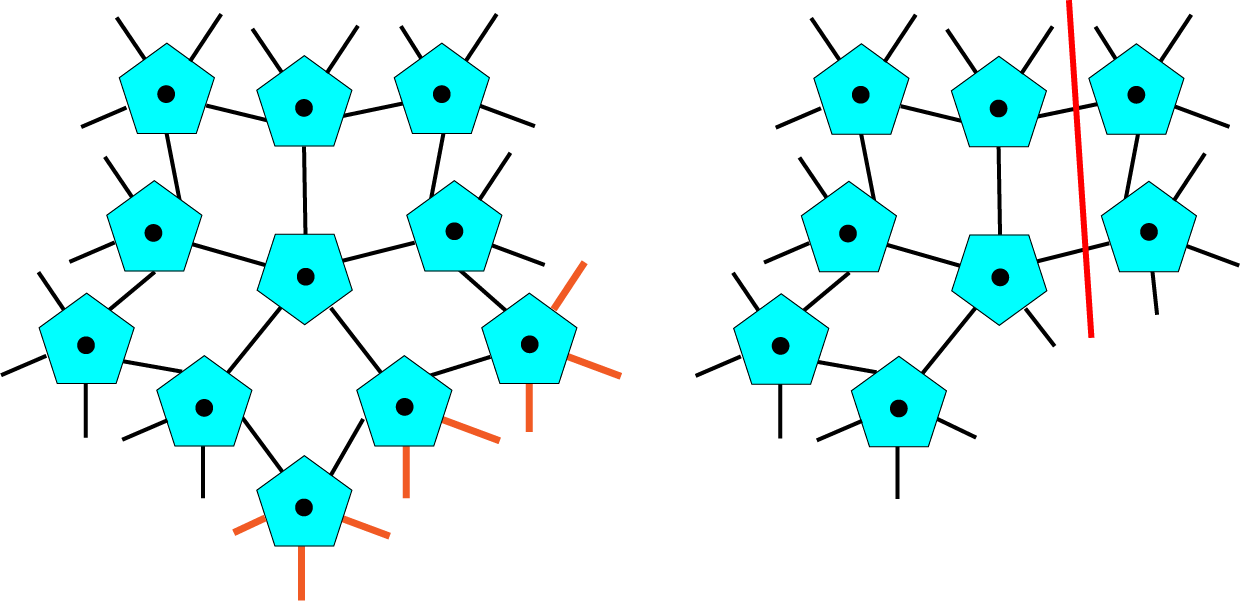}
    \caption{
      The process of tracing out boundary subregion $C$ (orange indices on the left) to arrive at
      an effective tensor network (right) without disturbing $\rho_{AB}$. The red line on the right is the area term for the effective tensor network, representing the entanglement wedge cross section.}
    \label{holographic_code}
\end{figure}

For the second setup, we investigate mixed states created by tracing out a subspace of an overall pure state (i.e. the bulk input state is pure). In doing so, we decompose the original error-correcting code into a tensor network that only has the physical degrees of freedom in boundary subsystems $A$ and $B$. In order to arrive at this effective tensor network, we must trace out the degrees of freedom of $C$ as seen in 
Fig.~\ref{holographic_code}. 
This involves removing all of the tensors in the entanglement wedge of $C$ by
repeatedly applying the Hermitian conjugates, $V^{\dagger}$, of the perfect
tensors in the entanglement wedge of $C$.
Once this process is completed,
we are left with a new tensor network 
with a simplified geometry. Here, we reach an impasse because
the new tensor network is no longer an isometry from the bulk logical indices to $A$ and $B$. 
We are then unable to repeat our argument from before
to find the negativity to be equivalent to the
entanglement wedge cross-section in holographic codes with the bulk quantum
correction. It would be interesting to better understand this effective tensor network in the context of quantum error correcting codes to derive a holographic negativity formula for tripartitions.

\section{Conjecture for AdS/CFT}

The QEC code considerations above suggest that the logarithmic negativity is captured by the minimal
entanglement wedge cross section.
We now need to address the differences between tensor networks and AdS/CFT.
For example, 
the spectrum of the entanglement Hamiltonian in holographic code models is
completely flat (i.e. $\ket{\chi}$ is maximally entangled),
while it is not in (holographic) CFTs.
This implies that in the full-fledged AdS/CFT
the area contribution in (\ref{main}) should describe some
backreacted geometry
analogous to the area contribution for the holographic duals of R\'enyi entropies \cite{2016NatCo...712472D,2019JHEP...05..052A,2019JHEP...10..240D}.
\subsection{Backreaction}

To address the issue of backreaction,
we briefly overview Dong's proposal for the holographic dual of R\'enyi entropy.
There, a close variant of the R\'enyi entropy is equal to the area of a cosmic brane with tension
\begin{align}
  n^2 \frac{\partial}{\partial n}
  \left( \frac{n-1}{n} S_n\right)
  = \frac{{\it Area}({\it \mbox{Cosmic Brane}}_n)}{4 G_N},
\end{align}
where $S_n$ is the $n^{th}$ R\'enyi entropy and cosmic branes are gravitating
objects living in the bulk.
The tension of the cosmic brane depends on the replica index as
\begin{align}
    T_n = \frac{n-1}{4 n G_N}.
    \label{tension_eq}
\end{align}
The cosmic brane is analogous to the RT surface except that it creates a conical deficit angle
\begin{align}
    \Delta \phi = 2 \pi \frac{n-1}{n}.
\end{align}
In order to find the corresponding backreacted geometry, one must find the classical solution to the equations of motion for the action
\begin{align}
  I = -\frac{1}{16\pi G_N}
  \int d^{d+1}x \sqrt{G}R + I_{{\it matter}}
  + I_{{\it brane}},
\end{align}
where
\begin{align}
    I_{brane} = T_n \int d^{d-1}y \sqrt{g}.
\end{align}
$G$ is the total bulk metric while $g$ is the induced metric on the brane. Note that the brane becomes tensionless in the replica limit ($n\rightarrow 1$), so the formula naturally reproduces the RT formula.

For negativity, we introduce backreaction in the bulk by defining a family of
area functions
in the ambient bulk of the entanglement wedge
\begin{align}
  \tilde{\mathcal{A}}_n \equiv n^2 \frac{\partial}{\partial n} \left( \frac{n-1}{n} \mathcal{A}_n\right)
  = \frac{{\it Area}({\it \mbox{Cosmic Brane}}_n)}{4 G_N},
  \label{brane_areas}
\end{align}
where the bulk gravitational solution now has boundary conditions on the boundaries of the entanglement wedge. We then naturally claim
\begin{align}
    \mathcal{E} = \lim_{n \rightarrow 1/2} {\mathcal{A}}_n + \mathcal{E}_{bulk}.
    \label{general_formula}
\end{align}
$\mathcal{A}_n$ is in general a very difficult problem to solve as one needs an analytic formula for $\tilde{\mathcal{A}}_n$. However, a special case of this is when the entangling surface is spherical, in which case, we know the effect of the backreaction. In this special case, the negativity is proportional to the tensionless brane ($n\rightarrow 1$) answer \cite{2011JHEP...12..047H,2014JHEP...10..060R}
\begin{align}
  \mathcal{E} = \mathcal{X}_d^{hol} \tilde{\mathcal{A}_1}+ \mathcal{E}_{bulk} = \mathcal{X}_d^{hol} \frac{E_W}{4 G_N}+ \mathcal{E}_{bulk},
  \label{sphere_neg}
\end{align}
where 
\begin{align}
      \mathcal{X}^{{\it hol}}_d = \frac{1}{2}x_d^{d-2}\left(1 + x_d^2\right) - 1, \label{coefficient} \\ \nonumber
  x_d = \frac{2}{d}
  \left(1 + \sqrt{1 - \frac{d}{2} + \frac{d^2}{4}}\right).  
\end{align}
We will use this simplification throughout this paper. Observe that when
$d=2$, $\mathcal{X}^{\textit{hol}}_2=3/2$ and  this relation is consistent with
(and follows alternatively from)
the fact that
$\mathscr{E}_A = S_{1/2}(\rho_A)$ and
$S_{n}(\rho_A)= (c/6)(1+1/n)\log\ell/\epsilon$
where $\ell$ is the radius (size) of the region (interval) $A$.
The coefficient
$\mathcal{X}^{{\it hol}}_d$
smoothly interpolates between the
$\mathcal{X}^{{\it hol}}_2 = 3/2$
that we will use for (1+1)d CFT's and $\mathcal{X}^{{\it hol}}_{\infty} = e-1 \sim 1.718$. Notably, for the $\mathcal{N} = 4$ SYM, $\mathcal{X}^{{\it hol}}_4 \sim 1.674$.

\subsection{Connection to entanglement of purification}
The minimal entanglement wedge cross section was studied
in \cite{2018NatPh..14..573U, 2018JHEP...01..098N}
as an interesting measure of entanglement in mixed states.
The authors of Refs.~\cite{2018NatPh..14..573U, 2018JHEP...01..098N} identified properties of this measure 
and matched these properties to a list of correlation measures in quantum
information theory.
They decided upon the entanglement of purification.
Entanglement of purification is a famously difficult quantity to obtain. 
It is also dependent on both quantum and classical correlations,
differing from the negativity which only measures quantum correlations.
Even so, the proposal for holographic EoP and our proposal for holographic
negativity do not contradict one another.
Rather, we identify
$\tilde{\mathcal{A}}_1$ in (\ref{brane_areas}) with the conjectured holographic EoP and note that the negativity and EoP
will be proportional only when the entangling surface is spherical and not for generic configurations.

It is also worth mentioning that
there is
yet another proposal for holographic negativity
which has been shown to produce the correct
behaviors for adjacent subsystems of (1+1)d CFT's \cite{2016arXiv160906609C, 2017arXiv170708293J}.
(There is also a higher-dimensional version of this proposal.)
The proposal relates the entanglement negativity in holographic CFTs to
a proper combination of the bulk minimal surface areas (geodesics).
For example, 
for the case of two adjacent intervals at zero temperature,
it was proposed that the entanglement negativity is given by
\begin{equation}
  \mathscr{E} = \frac{3}{16 G_N} (\mathscr{L}_{A_1} + \mathscr{L}_{A_2} - \mathscr{L}_{A_1, A_2}),
\label{IIT}
\end{equation}
where $\mathscr{L}_{A_1,A_2}$ is the area of the codimension-2 extremal surface homologous to the union of $A_1$ and $A_2$.
As a corollary of this conjectured formula,
the holographic negativity is related to the mutual information of the two
intervals as  
\begin{equation}
\mathscr{E} = \frac{3}{4} {I}(A_1, A_2).
\end{equation}

Overall, there seems to be an intriguing connection between three quantum information theoretical quantities
in holographic theories:
the entanglement negativity,
the entanglement of purification,
and
the mutual information.

Unfortunately, computing the entanglement of purification would be rather difficult
in general.
Using random stabilizer tensor networks \cite{2020PhRvL.125x1602N},
we can compare the three quantities of interest: entanglement negativity, entanglement of purification, and mutual information. We look at a tripartition of the boundary. It was stated in Ref.~\cite{2018JHEP...01..098N} that any such tripartition can be decomposed into Bell and GHZ-like states, up to unentangled states:
\begin{align}
&  U_A U_B U_C \ket{\Psi}_{ABC} 
  \\ \nonumber
  &\quad 
    =(\ket{\Phi}_{A_1 B_1})^c
    (\ket{\Phi}_{B_2 C_1})^a(\ket{\Phi}_{A_2 C_2})^b
    (\ket{{\it GHZ}}_{A_3 B_3 C_3})^g, 
\end{align}
with
\begin{align}
  &
  \ket{\Phi}_{A B}
  = \frac{1}{\sqrt{p}}\sum_{i=0}^{p-1}\ket{i}_A \ket{i}_B,
  \nonumber \\
  &  \ket{{\it GHZ}}_{A BC}
  = \frac{1}{\sqrt{p}}\sum_{i=0}^{p-1}\ket{i}_A \ket{i}_B \ket{i}_C.
\end{align}
It is straightforward to then show that the negativity {equals $c \log p$, the} entanglement of
purification equals $(c + g) \log{p}$, and half the mutual
information equals $(c+ \frac{g}{2}) \log{p}$.
All three of these are coincident in the large-$N$ limit of \cite{2020PhRvL.125x1602N},
$E_P(A,B) =\mathscr{E}(A,B)= (1/2)I(A,B)$,
which is the standard limit when dealing with random tensor networks.

\section{Entanglement wedge and negativity in \texorpdfstring{$AdS_3/CFT_2$}{TEXT}}
\label{Entanglement wedge and negativity in AdS3CFT2}

In the following,
we will make more detailed comparisons
between
the entanglement negativity
and the minimal entanglement wedge cross section
in the context of ${\it AdS}_3/{\it CFT}_2$:
Specifically,
the entanglement negativity here will be computed
using the properties of holographic ${\it CFT}_2$.
When possible, we also compare these with
(suitable linear combinations of) the mutual information. 

Paralleling the discussion in tensor networks, 
we are interested in the following basic set ups:
\begin{enumerate}
\item The case of single interval:
  In this case, we bipartition the total
  space into a single interval $A$ and its compliment $A^c$,
  and consider the entanglement negativity $\mathscr{E}^A$.
  The system can be in its ground state or in more generic
  pure or mixed states. However, our main focus will be
  cases of mixed states, in particular, the system at finite
  temperature, since for pure states the entanglement negativity
  is simply the R\'enyi entropy with R\'enyi index $1/2$ and our conjectured holographic formula goes through simply.
  
\item 
  The case of two intervals:
  In this case, we start from the ground state
  and tripartition the total system
  into intervals $A_1, A_2$ and $B$.
  We trace out $B$ and discuss the entanglement
  negativity of the reduced density matrix $\rho_{A_1 A_2}$
  for the two intervals $A_{1,2}$. The two intervals can be right next to each other (adjacent)
  or can be separated (disjoint) by the interval $B$.
\item
  In addition, we will consider the
  entanglement negativity of the thermofield double state;
  Here, we take the partial transpose in either one of
  the Hilbert spaces, and discuss the entanglement negativity.
  
\end{enumerate}

Let us warm up by considering a single-interval at zero temperature. As previously mentioned, the entanglement negativity of the interval is equal to the R\'enyi entropy at R\'enyi index 1/2. For 1+1d CFT's, the R\'enyi entropies are simply determined by the central charge
\begin{equation}
   S_n = \frac{c}{12} \left(1 + \frac{1}{n}
     \right)\log\left(\frac{\ell}{\epsilon}\right),
\end{equation}
where $\ell$ is the length of the interval and $\epsilon$ is a UV cutoff, and hence in this case the negativity is given by 
\begin{align}
  \label{E and S1/2}
  \mathcal{E} = S_{1/2} = \frac{c}{4} \log\frac{\ell}{\epsilon}.
\end{align}
Noting that the minimal entanglement wedge cross section in this case is equal to the length of the RT surface (= the von Neumann entanglement entropy), we confirm that
\begin{align}
  \mathcal{E} = \frac{3}{2}E_W.
\end{align}

\subsection{Two intervals}

\subsubsection{Adjacent intervals}
\label{Adjacent inervals}

We start with the entanglement negativity
at zero temperature for two intervals $A_{1,2}$, which can be adjacent or disjoint.
Our starting point is 
the expression of 
the moment $\mathrm{Tr}\, (\rho^{T_2})^{n_e}$
as a correlation function of the twist operators
\cite{2012PhRvL.109m0502C,2013JSMTE..02..008C}:
\begin{align}
  \label{negativity and twist op}
  \mathscr{E}
  &= \lim_{n_e \to 1}
  \log\mathrm{Tr}\, (\rho^{T_2})^{n_e}
    \nonumber  \\
  &=
    \lim_{n_e \to 1} \log
  \langle \sigma_{n_e}(w_1, \bar{w}_1) \bar{\sigma}_{n_e}(w_2, \bar{w}_2)
    \nonumber \\
  &\qquad \qquad
    \times 
  \bar{\sigma}_{n_e}(w_3, \bar{w}_3) \sigma_{n_e}(w_4, \bar{w}_4)
  \rangle_{\mathbb{C}}.
\end{align}
Here, the conformal dimension of
the twist operator $\sigma_n$ is 
\begin{align}
  h_n = \frac{c}{24} \left(n - \frac{1}{n}\right).
\end{align}
The complex Euclidean coordinates
$w= i \tau +x$
are set to be
$w_1=Y_1$,
$w_2=Y_2$,
$w_3=X_1$,
$w_4=X_2$,
with
\begin{align}
  X_1 - X_2 &= \ell_1,
              \quad
              Y_1 - Y_2 = \ell_2,
              \quad
              Y_2 - X_1 = d,
\end{align}
where $\ell_{1,2}$ is the length of the interval $A_{1,2}$
and $d$ is the distance between the intervals.

In the limit of the adjacent intervals, $d\to 0$,
the negativity is given by the three-point function, 
\begin{align}
  \mathscr{E}
  &=
    \lim_{n_e \to 1} \log
  \langle \sigma_{n_e}(w_1, \bar{w}_1) \bar{\sigma}^2_{n_e}(w_2, \bar{w}_2)
  \sigma_{n_e}(w_4, \bar{w}_4)
  \rangle_{\mathbb{C}},
\end{align}
and hence completely determined by conformal symmetry. 
Using the dimension of the twist operator, 
one then obtains
\begin{align} \label{neg cft}
  \mathscr{E} = \frac{c}{4}\log\left[\frac{\ell_1 \ell_2}{\ell_1 + \ell_2} \right]
  + {\it const.}
\end{align}

Let us compare the negativity \eqref{neg cft}
with the minimal cross section
of the corresponding entanglement wedge,
which is given, according to Ref.~\cite{2018NatPh..14..573U}, by:
\begin{align}
  E_W
  &=
    \left\{
    \begin{array}{ll}
      \displaystyle
    \frac{c}{6}
      \log\frac{1+\sqrt{x}}{1-\sqrt{x}},
      &
      \displaystyle
        \quad
       \frac{1}{2} \le x \le 1 
        \\
  \\
    0,
      &
      \displaystyle
        \quad
      0\le  x\le \frac{1}{2}
      \end{array}
    \right.
\end{align}
where $x$ is the cross ratio,  
\begin{align}
  x := \frac{w_{12}w_{34}}{ w_{13} w_{24}}
  =
  \frac{\ell_1 \ell_2}{(\ell_1 +d)(\ell_2 +d)}.
\end{align}
In the limit of adjacent intervals $d\to 0$,
\begin{align}
  \label{min ent wedge cross section}
  E_W
  &\to
    \frac{c}{6} \log\left( 4z \right)
  = 
    \frac{c}{6} \log\left(
    \frac{4}{\epsilon}
    \frac{\ell_1 \ell_2}{ (\ell_1 +\ell_2)}
    \right).
\end{align}
Thus, if the constant in \eqref{neg cft} is properly chosen, 
$\mathscr{E} = (3/2) E_W$.

Let us also consider the following, properly normalized,
mutual information for the two intervals:
$
(3/4) I(A_1, A_2)
$.
This claim follows from the proposed holographic formula
for the entanglement negativity
for the mixed state of the adjacent intervals
\begin{align}
 \mathscr{E} 
  =
  \frac{3}{4}
  \cdot 
  \frac{1}{4G_N}
  \left[
  \mathscr{L}_{12}+\mathscr{L}_{23}
  -
  \mathscr{L}_{13}
  \right]
  \label{MI_neg_formula}
\end{align}
where $\mathscr{L}_{12}$ etc.~are the bulk geodesic lengths.
It is straightforward to check that 
$(3/4) I(A_1, A_2)$
is also given by
$
(c/4) \log[ \ell_1 \ell_2/(\ell_1+\ell_2)] +
{\it const}
$.
Summarizing, for adjacent intervals, all the three quantities are equal,
\begin{align}
  \mathscr{E} = \frac{3}{2}E_W = \frac{3}{4} I(A_1,A_2).
\end{align}

We now generalize to a thermal state. We take adjacent intervals of equal length $\ell$. 
For finite temperature, the negativity of adjacent intervals is computed by the
following three-point
function of twist fields on the cylinder
\begin{equation}
    \mathcal{E} = \lim_{n_e \rightarrow 1} \log\left( \langle \sigma_{n_e}(z_1) \bar{\sigma}_{n_e}^{2}(z_2)\sigma_{n_e}(z_3) \rangle_{\beta} \right).
\end{equation}
Unlike the case for thermal bipartite negativity, there are no ambiguities regarding transforming from the complex plane to the cylinder. This is due to the adjacent intervals being finite \cite{2015JPhA...48a5006C}. We use the following map from the complex plane to the cylinder
\begin{align}
  w(z) &= e^{2 \pi z/ \beta},
  \\ \nonumber
  \mathcal{E} &= \left(\frac{2 \pi}{\beta} \right)^{-c/4}\langle \sigma(e^{-2\pi \ell/\beta})
                \bar{\sigma}^2 (1) \sigma(e^{2\pi \ell/\beta}) \rangle_{\mathbb{C}},
    \label{conf_trans}
\end{align}
where we have taken the replica limit. We then compute the three-point function to arrive at a negativity of 
\begin{equation}
    \mathcal{E} = \frac{c}{4}\log\left[ \frac{\beta}{2 \pi \epsilon} \tanh \left(\frac{\pi \ell}{\beta} \right) \right],
\end{equation}
where we have introduced the regulator $\epsilon$.

We can now do the corresponding calculation holographically. We use the planar BTZ geometry 
\begin{equation}
    ds^2 = -\frac{(r^2 - r_H^2)}{R^2}dt^2 + \frac{dr^2}{r^2 - r_H^2} + \frac{r^2}{R^2}dx^2.
\end{equation}
Due to the symmetry of the setup, the minimal cross-section is purely radial
\begin{equation}
    \Sigma = \int_{r_*}^{r_{\infty}} \frac{dr }{\sqrt{r^2 - r_H^2}},
\end{equation}
where $r_*$ is the location of the turning point which is related to the interval length by
\begin{equation}
    r_* = r_H \coth(\ell r_H).
\end{equation}
Using (\ref{main}), we arrive at
\begin{equation}
    \mathcal{E} = \frac{3}{8 G_N} \Sigma = \frac{c}{4}\log\left[ \frac{\beta}{2 \pi \epsilon} \tanh \left(\frac{\pi \ell}{\beta} \right) \right],
\end{equation}
which exactly matches the CFT result. We note that the same answer has been found using (\ref{MI_neg_formula}) \cite{2017arXiv170708293J}.

\subsubsection{Disjoint intervals}
\label{Disjoint intervals}

While the negativity for adjacent intervals is given in terms of the three-point function and hence universal, the negativity for disjoint intervals depends on the full operator content of the theory. Let us examine in the case of the holographic CFT in the large-$c$ limit, using the result from Ref.~\cite{2014JHEP...09..010K}.  

Starting from \eqref{negativity and twist op}, using a conformal map that sends $w_1 \to \infty$, $w_2 \to 1$, $w_3 \to x$, and $w_4 \to 0$, the negativity is written as 
\begin{align}
  \label{neg disjoint}
  \mathscr{E}
    &=
      \lim_{n_e\to 1}
                     \log(|w_{24}|^{-4h_{n_e}}  |w_{13}|^{-4 h_{n_e}})
      \\ \nonumber
  &\quad
      +
    \lim_{n_e \to 1}
                     \log
                     \left[ 
                     \langle
    \sigma_{n_e}(\infty) \bar{\sigma}_{n_e}(1)
    \bar{\sigma}_{n_e}(x,\bar{x}) \sigma_{n_e}(0)
                     \rangle
                     \right].
\end{align}
The first term does not contribute in the replica limit since $h_{n_e}\to 0$. Hence, the sole contribution (the second term in \eqref{neg disjoint}) depends only on $x$ and the negativity for two disjoint intervals (at zero temperature and for infinite systems) is a scale invariant quantity.

We now try to find behaviors of the
universal function
$\langle \sigma_{n_e}(\infty) \bar{\sigma}_{n_e}(1)$
$\bar{\sigma}_{n_e}(x,\bar{x}) \sigma_{n_e}(0) \rangle$
in the large-$c$ limit. 
It can be expanded in terms of
the conformal blocks as
\begin{align}
  &
\langle \sigma_{n_e}(\infty) \bar{\sigma}_{n_e}(1)
    \bar{\sigma}_{n_e}(x,\bar{x}) \sigma_{n_e}(0) \rangle
  \nonumber \\
  &\quad 
  =
  \sum_p a_p\, 
  \mathcal{F}(c, h_p, h_i, x)
  \bar{\mathcal{F}}(c, \bar{h}_p, \bar{h}_i, \bar{x}),
\end{align}
where $p$ labels operators in intermediate OPE channels
with conformal dimension $h_p$;
$h_i$ collectively represents
the conformal dimensions of four ``external'' operators,
i.e., $h_{n_e}$;
$a_p$ is a constant depending on the OPE coefficients.
In the decomposition of the conformal block
we assume there is a single dominant channel $p$,
and disregard other contributions \cite{2013arXiv1303.6955H}
\begin{align}
  &
\langle \sigma_{n_e}(\infty) \bar{\sigma}_{n_e}(1)
    \bar{\sigma}_{n_e}(x,\bar{x}) \sigma_{n_e}(0) \rangle
  \nonumber \\
  &\quad
  \sim 
   \mathcal{F}(c, h_p, h_i, x)
  \bar{\mathcal{F}}(c, \bar{h}_p, \bar{h}_i, \bar{x}).
\end{align}
For large-$c$ CFTs,
the conformal block exponentiates as \cite{1987TMP....73.1088Z}
\begin{align}
   \mathcal{F}(c, h_p, h_i, x)
  \sim 
   \exp \left[
   -\frac{c}{6} f\left( \frac{h_p}{c}, \frac{h_i}{c}, x \right)
  \right].
\end{align}
Hence, 
assuming $f( {h_p}/{c}, {h_i}/{c}, x ) =f( {\bar{h}_p}/{c}, {\bar{h}_i}/{c}, \bar{x} )$,
\begin{align}
  &
  \log
\langle \sigma_{n_e}(\infty) \bar{\sigma}_{n_e}(1)
    \bar{\sigma}_{n_e}(x,\bar{x}) \sigma_{n_e}(0) \rangle
   \sim
   -\frac{c}{3} f\left( \frac{h_p}{c}, \frac{h_i}{c}, x \right).
\end{align}

In Ref.~\cite{2014JHEP...09..010K},
the dominant channel when $x\to 1$
(the limit of adjacent intervals $d/\ell_{1,2}\to 0$)
is identified
as the double twist operator $\sigma^2_{n_e}$
with conformal dimension
$h_p = h^{(2)}_{n_e} =
(c/12) (n_e/2 - 2/n_e)$.
On the other hand,
when $x\to 0$
(the limit where the distance between of two intervals
is large $d/\ell_{1,2}\to \infty$),
the dominant channel is vacuum.
The analysis in the latter case
($x\to 0$) is
similar (identical) to the
case of the entanglement entropy
of two disjoint intervals;
it is exponentially small.
(For small $x$, the computation of the four-point function is identical to the one performed for
entanglement entropy, and there is a factor of $(n-1)$
which vanishes in the $n\to 1$ limit.)
In the following, we will mainly focus on
the case of $x\to 1$.

\paragraph{Monodromy method}

The function $f$ can be found by using the monodromy method
and this program was carried out in Ref.~\cite{2014JHEP...09..010K}.
The same kind of approximation was used
to compute the mutual information
for disjoint intervals 
in holographic CFT in \cite{2013arXiv1303.6955H}
to reproduce the result from the RT formula.
There, as the distance between the two intervals
increases/decreases, there is a ``phase transition''
and the mutual information has a ``singularity''
as a function of the distance between the intervals \cite{2010PhRvD..82l6010H}.
We expect there is a
similar phase transition in the entanglement negativity
\cite{2014JHEP...09..010K}.

In the monodromy method, 
the large-$c$ conformal block
$f$ is given in terms of
the accessory parameter
$c_2$ as $\partial f/\partial x = c_2(x)$.
In Ref.~\cite{2014JHEP...09..010K},
two solutions were found numerically in the monodromy problem,
which are approximately given by 
\begin{align}
  \label{accessory parameter pm}
  &y(1-y)c^{\pm}_2(1-y) = -\frac{3}{4} + \frac{3}{4}
                   \left( \frac{1}{2}\pm \frac{1}{4} \right)
                   y + \cdots,
                   \\ \nonumber
  &c^-_2 (x) \sim \frac{3 (x+3)}{16 x (x-1) },
              \quad
  c^+_2 (x) \sim \frac{3 (3x+1)}{16 x (x-1) },
\end{align}
where $y=1-x$.
By integrating $c_2$, 
these accessory parameters are translated to conformal blocks 
in the vicinity of $x=1$,
\begin{align}
  f^-(x) &=
  \frac{3}{16}
    \log\left[
    \frac{(1 - x)^4}{x^3} \right]
         +{\it const.},
         \\ \nonumber
  f^+(x) &=
           \frac{3}{16}
           \log\left[ \frac{(1-x)^4}{x} \right]
         +{\it const.}
\end{align}
If the dominant solution ($f^-(x)$ in this case) is chosen,
the entanglement negativity is given by
\begin{align}
  \mathscr{E}^-
  &\sim
  -\frac{c}{3}
  \cdot
  \frac{3}{16}
    \log
    \left[
    \frac{(1 - x)^{4}}{x^{3}} 
  \right]
  +{\it const.}
\end{align}
As in the case of adjacent intervals,
the constant has to be chosen properly,
which cannot be determined from the monodromy method.
We set 
\begin{align}\label{const}
  {\it const.} =
  \frac{c}{4} \log(4).
\end{align}
(See
 \eqref{neg cft} and \eqref{min ent wedge cross section}
in Sec.~\ref{Adjacent inervals}.) 

\begin{figure}[t]
  \begin{center}
  \includegraphics[width = 7cm]{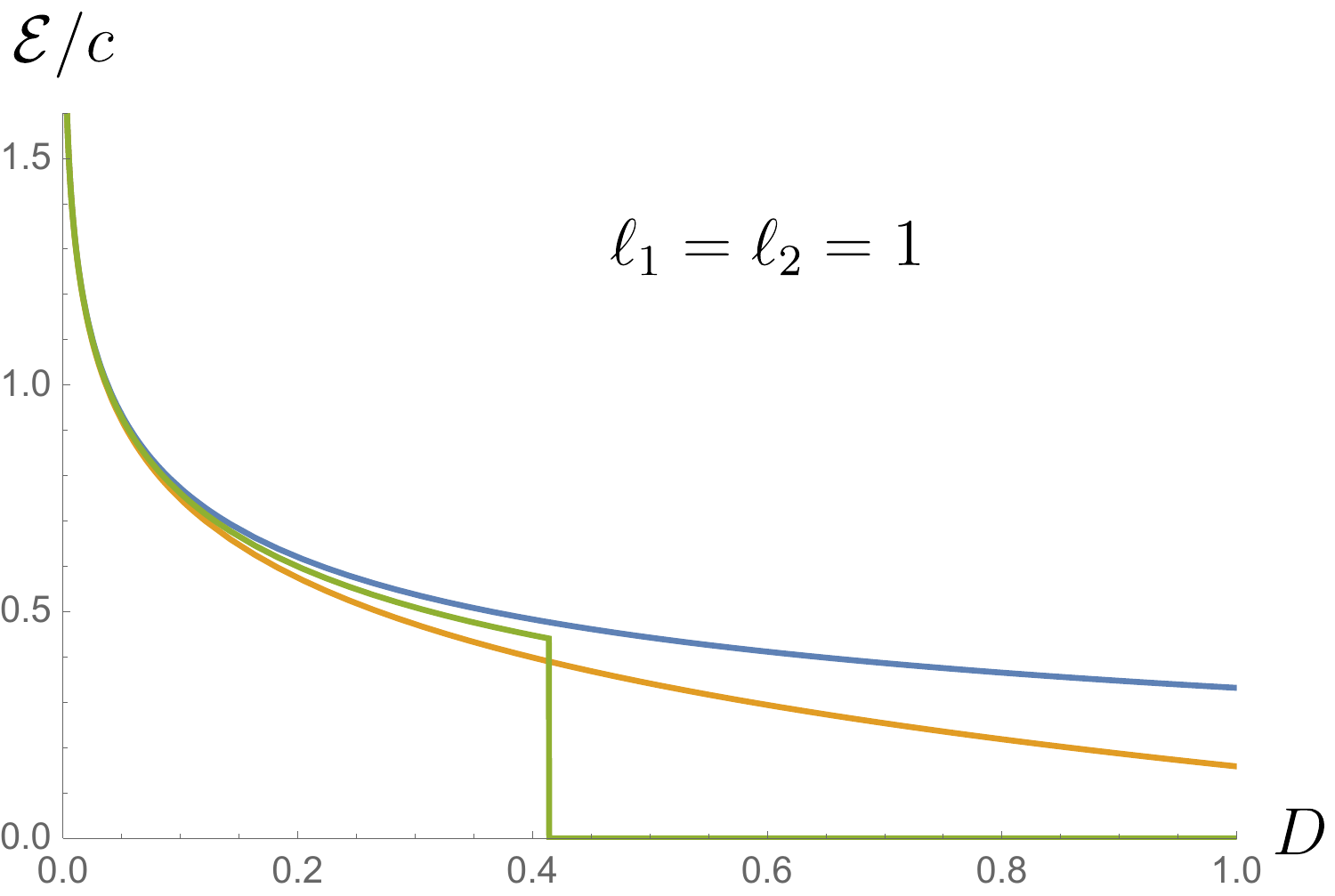}
  \end{center}
  \caption{
    \label{neg_disjoint}
    (Blue and yellow)
    The logarithmic negativity for disjoint intervals
    at zero temperature 
    for holographic CFT computed
    from the large-$c$ conformal blocks $f^{\pm}(x)$
    as a function of the distance $d$
    between the two intervals.
    (Green)
    The minimal entanglement wedge cross section,
    $(3/2)E_W$
    plotted in the unit of $c$.
  }
\end{figure}

The entanglement negativity,
computed by using the solutions $f^{\pm}$,
are plotted in Fig.~\ref{neg_disjoint},
together with the minimal entanglement wedge cross section.
Note that the above solutions are valid for $x\sim 1$.
On the other hand,
for sufficiently small $x$
there is a phase transition to the other branch,
where negativity is simply zero.
While
the entanglement negativity
$\mathcal{E}^{\pm}$ and the minimal entanglement wedge cross section
$(3/2)E_W$ disagree,
it is interesting to note
that
the minimal entanglement wedge cross section
is right in between
the two solutions.

To have a closer comparison with the
minimal entanglement wedge cross section $(3/2)E_W$,
\eqref{min ent wedge cross section},
we define an analogue of conformal block $\mathcal{F}^W(x)$ by
\begin{align}
  \frac{3}{2}E_W(x)
  =:
  \log
  \left[a_p\,  \mathcal{F}^W(x) \bar{\mathcal{F}}^W(x) \right].
\end{align}
Choosing $a_p=4^{c/4}$
(i.e., $\log a_p = (c/4)\log(4)$;
See \eqref{const}.), 
\begin{align}
  \mathcal{F}^W(x)
  &=
  \left[
    \frac{1}{4}
  \frac{1+\sqrt{x}}{1-\sqrt{x}}
  \right]^{\frac{c}{8}}.
\end{align}
$\mathcal{F}^W(x)$ can be expanded in small $y=1-x$
as
\begin{align}
  \mathcal{F}^W(x)
  &=
    y^{-\frac{c}{8}}
    \Bigg[
    1-\frac{c y}{16}
    +\frac{c(c-12)}{512} y^2
    \nonumber \\
  &\quad 
    -\frac{c \left(c^2-36 c+320\right)}{24576} y^3
    +\cdots
    \Bigg].
\end{align}
Further introducing 
the corresponding accessory parameter by 
\begin{align}
  c^W_{2}(x) &:=
    -
              \frac{3}{4}
              \frac{6}{c}
              \frac{d E_W}{dx}
  =
    -
  \frac{3}{4}
  \frac{1}{(1-x)\sqrt{x}},
\end{align}
we see that $c^-_2$ and $c^W_{2}$ disagree at linear order in $y$:
\begin{align}
  \label{expansion c2w}
  y (1-y)c^W_{2}(1-y) &= 
                       -\frac{3}{4} +\frac{3}{8} y + 
                       \frac{3}{32}y^2 +\cdots.
\end{align}

\paragraph{Series expansion}

Ref.~\cite{2014JHEP...09..010K}
also looked at the expansion of the conformal
block in terms of the cross-ratio:
\begin{align}
  \label{exp}
  &
  \mathcal{F}(h_p, y)
  =
  y^{h_p}\Big[
  1 + \frac{h_p}{2} y
  + 
  \frac{ h_p(h_p+1)^2}{4 (2h_p+1)} y^2
    \nonumber \\
  &+
  \frac{h^2_p (1-h_p)^2}{ 2 (2h_p+1) [ c(2h_p+1) + 2h_p (8h_p-5)]} y^2
  +\cdots
  \Big],
\end{align}
where once again $y=1-x$.
Setting $h_p = -c/8$, 
we obtain 
\begin{align}
  \mathcal{F}(h_p, y)
  =
  y^{-\frac{c}{8}}\left[
  1
  -
  \frac{cy}{16}
  +
  \frac{(c-16)cy^2}{576} 
  +\cdots
  \right].
\end{align}
This is supposed to be valid for any $c$,
but as we will see, there is a complication.  
It seems that
the $h_p\to -c/8$ limit and the large-$c$ limit do not commute.

One reason is that, 
for generic values of $h_p$,
the third term is of order $(cy)^2$
and the forth term is of order $c y^2$,
while when $h_p= -c/8$,
they are both of the same order.
On the other hand, 
for $h_p\sim ac$ with $c$ large and generic value of $a\neq -1/8$,
we keep leading order terms $(cy)^n$.
For example, 
in the above expression \eqref{exp},
the third term is of order $(cy)^2$,
while 
the last term is sub leading as
$\sim cy^2$.
Collecting the $(cy)^n$ terms, 
\begin{align}
  \label{coeff}
  \mathcal{F}(h_p, y)
  =
  y^{h_p}\left[
  1 + \frac{h_p}{2} y
  +
  \frac{ h_p^2}{8} y^2
  +
  \frac{ h^3_p}{48} y^3
  +\cdots
  \right].
\end{align}
On the other hand,
from the entanglement wedge cross section,
keeping leading order terms,
\begin{align}
    \label{entanglement wedge leading}
  \mathcal{F}^W(x)
  &=
    y^{-\frac{c}{8}}
    \left[
    1-\frac{c y}{16}
    +\frac{c^2y^2}{512} 
    -\frac{c^3 y^3}{24576} 
    +\cdots
    \right].
\end{align}
Substituting $a=-1/8$ in \eqref{coeff},
\eqref{coeff} matches precisely with 
\eqref{entanglement wedge leading}.
Note also that \eqref{coeff}
can be exponentiated as
$  {\cal F}(h_p, y)
  =
  \exp
  \left[
  -
  ({c}/{6})
  f(x)
  \right]
$
with
\begin{align}
  f(x) &= -\frac{6}{c}
  \left[
   c \left(
  a \log y + \frac{a}{2}y
  +\cdots
  \right) \right].
\end{align}
The corresponding accessory parameter is given by
$
  c_2(x) ={\partial f}/{\partial x}
           =
           3 a + {6 a}/({1-x})
$.
Expanded in $y$ and substituting $a=-1/8$ naively,
\begin{align}
  y (1-y) c_2(1-y) &= -\frac{3}{4} + \frac{3}{8} y.
\end{align}
This expansion matches with
the expansion of the entanglement wedge cross section $c^W_{2}$,
\eqref{expansion c2w}.
This is consistent with the
result from the monodromy method.
However, of course, $a= -1/8$ is precisely
the point where various complications arise,
as seen from \eqref{exp}:
some of the expansion coefficients in \eqref{exp} diverge.

\subsection{Single interval at finite temperature}

Let us now discuss the case of single interval at finite temperature.
In this case, the negativity can be expressed as
\cite{2015JPhA...48a5006C}
\begin{align}
 \mathscr{E} = \lim_{L\to \infty}\lim_{n_e \to 1}
  \log\left[
  \langle
  \sigma_{n_e}(-L)
  \bar{\sigma}^2_{n_e}(-\ell)
  \sigma^2_{n_e}(0)
  \bar{\sigma}_{n_e}(L)
  \rangle_{\beta}
  \right],
\end{align}
where the conformal dimensions of $\sigma_{n_e}$ and $\sigma^2_{n_e}$ are given
by
\begin{align}
  h_{n_e} &=
  \frac{c}{24} \left(  n_e - \frac{1}{n_e} \right),
            \nonumber \\
  h^{(2)}_{n_e} &= 2 h_{\frac{n_e}{2}} = \frac{c}{12} \left(
  \frac{n_e}{2} - \frac{2}{n_e}
  \right).
\end{align}
Here, the order of the limits
is important; the replica limit has to be taken
before the $L\to \infty$ limit.
Below, we use the twist operator
formula to compute the entanglement negativity. 
Noting $h_n \to 0$ and $h^{(2)}_n \to - c/8$
in the replica limit,
the negativity is given by
\begin{align}
  &
  \mathscr{E}
  = \frac{c}{2}
    \log\left[
    \frac{\beta}{2\pi}
    e^{\frac{ \pi \ell }{ \beta}}
    \right]
  \nonumber \\
  &\quad 
  +
    \lim_{L\to \infty}\lim_{n_e \to 1}
    \log
    \langle
    \sigma_n(\infty) \bar{\sigma}^2_n(1)
    \sigma^2_n(x,\bar{x}) \bar{\sigma}_n(0)
    \rangle.
\end{align}
Here, the cross ratio in the $L\to \infty$ limit is 
\begin{align}
  x
  &=
 \frac{\big(1-e^{\frac{2 \pi  L}{\beta }}\big)
  \big(e^{-\frac{2 \pi  L}{\beta }}-e^{-\frac{2 \pi  \ell}{\beta }}\big)}
  {\big(e^{-\frac{2 \pi L}{\beta }}-1\big)
  \big(e^{-\frac{2 \pi \ell}{\beta }}-e^{\frac{2 \pi  L}{\beta}}\big)}
      \to
  e^{-2\pi \ell/\beta}.
\end{align}
As for the conformal block part, 
one can derive its semiclassical approximation by
using the monodromy method
\begin{align}
& \log\langle
   \sigma_n(\infty) \bar{\sigma}^2_n(1)
   \sigma^2_n(x,\bar{x}) \bar{\sigma}_n(0)
\rangle
                =
                -\frac{c}{3} f(x),
\end{align}
where $f(x)$ can be computed for $x\sim 0$ ($s$-channel)
and $x\sim 1$ ($t$-channel) separately
by using the monodromy method.

\subsubsection{\texorpdfstring{$t$}{TEXT}-channel}

In this channel, the identity block is dominant. 
The monodromy calculation is straightforward and
gives
\begin{align} \label{t-channel sol}
  f(x) &= \frac{12 h^{(2)}_{n_e}}{c} \log(1-x), \quad x \to 1.
\end{align}
This is a very similar situation as the entanglement entropy;
the vacuum (identity) block is completely determined
by the primary, and no effects from descendants.
Recalling $x\to e^{-2\pi \ell/\beta}$ in $L\to \infty$,
\begin{align} \label{t-channel neg}
  \mathscr{E}
  &=
    \frac{c}{2} \log\frac{\beta}{2\pi}
    +
           \frac{c}{2}
           \log
           \left(
           2 \sinh \frac{\pi \ell}{\beta}
           \right),
       \quad x \to 1.
\end{align}
Note that the proper cut off factor is missing in these expressions.
We simply replace $\beta/2\pi \to \beta/ (2\pi \epsilon)$.

\begin{figure}[t]
  \begin{center}
  \includegraphics[width = .8 \columnwidth]{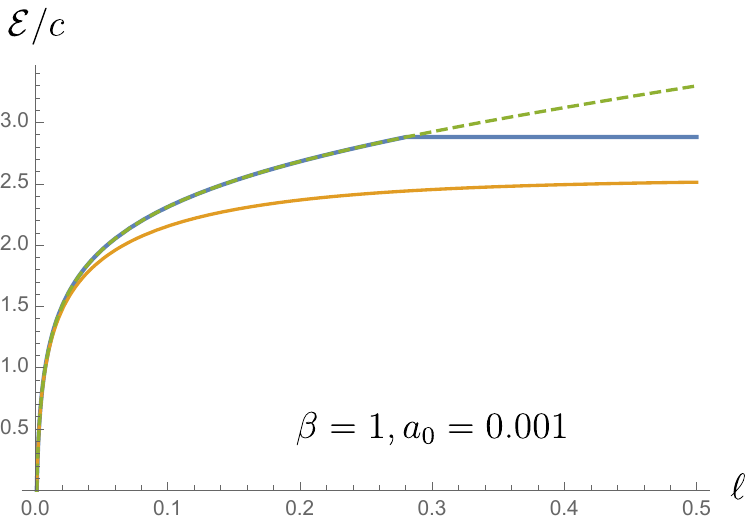}
  \end{center}
  \caption{The logarithmic negativity for a interval of length $\ell$ at finite temperature $\beta$ for holographic CFT. (Yellow) the generic CFT result (\ref{generic_CFT_bip}) with $g=0$. (Green) The negativity \eqref{t-channel neg} computed from the $t$-channel solution \eqref{t-channel sol} for $x\sim 1$. (Blue) the minimal entanglement wedge cross section. \label{single_finiteT}
  }
\end{figure}

The above result can be compared with
the minimal entanglement wedge cross section
\cite{2018NatPh..14..573U},
\begin{align}
  \label{entanglement wedge finite T}
  \frac{3}{2} E_W
  =
  \frac{c}{2}
  \mbox{min}\, 
  \left[
  \log
  \left(
  \frac{\beta}{\pi \epsilon}
  \sinh \frac{\pi \ell}{\beta}
  \right),
  \, 
  \log
  \left(
  \frac{\beta}{\pi \epsilon}
  \right)
  \right]
\end{align}
and the generic CFT result \cite{2015JPhA...48a5006C}
\begin{align}
  \mathscr{E}=
  \frac{c}{2} 
  \log
  \left(
  \frac{\beta}{\pi a_0}
  \sinh \frac{\pi \ell}{\beta}
  \right)
  -
 \frac{\pi c \ell}{2\beta}
  + g( e^{- 2\pi \ell/\beta})
  \label{generic_CFT_bip}
\end{align}
with $g=0$.
With the choice $g=0$,
this is the result predicted by
the holographic negativity proposal of
Ref.~\cite{2016arXiv160906609C, 2017arXiv170708293J}. However, as we just found, it is not valid to set $g = 0$ generically because the full operator content is important at leading order. Setting $g = 0$ is is only a good approximation near $x = 1$. 
Near $x =1$, all three quantities
(the entanglement negativity,
the minimal entanglement wedge cross section,
and the mutual information) agree (Fig.~\ref{single_finiteT}).

\subsubsection{\texorpdfstring{$s$}{TEXT}-channel}

In this channel, the dominant operator
is the twist operator $\sigma_n$ with dimension $h_n$.
The semiclassical conformal block can be obtained
by solving the monodromy problem around
$(x,0)$ with the trivial monodromy
$\mathrm{Tr}\, M_{(x,0)}=2$ in the replica limit.

\begin{figure}[t]
  \begin{center}
  \includegraphics[width = .8 \columnwidth]{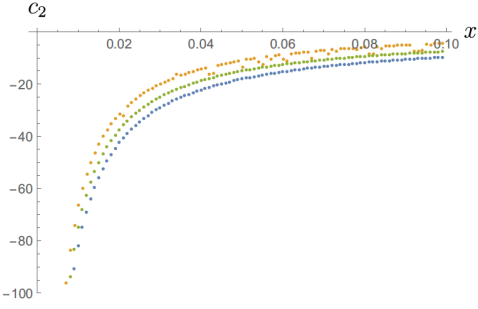}
  \end{center}
  \caption{
    The accessory parameter $c_2$ as a function of
    the cross ratio $x$.
    There are two solutions (Blue and Yellow)
    centered around
    $c^W_2(x)= -3/(4x)$ (Green).
    \label{c2_single_interval}
  }
\end{figure}

Numerical solutions of the monodromy problem
are shown in Fig.~\ref{c2_single_interval}.
The situation here is similar to
the negativity for disjoint intervals for $x\sim 1$.
(See Sec.~\ref{Disjoint intervals}, around \eqref{accessory parameter pm}.)
There are two solutions $c^{\pm}_2$ centered around
\begin{align}
  c^W_2 (x) = -\frac{3}{4}\frac{1}{x}.
\end{align}
The accessory parameter $c^W_2(x)$ is 
consistent with the minimal entanglement wedge cross section
\eqref{entanglement wedge finite T}:
The corresponding conformal block, up to a unknown constant,
is given by
\begin{align}
  f^W(x) &= \frac{6 h^{(2)}_{n_e}}{c}\log(x) = -\frac{3}{4}\log x, \quad x\rightarrow 0, \nonumber \\
  \mathcal{F}^W(x)
  &= x^{\frac{c}{8}}.
\end{align}
If we use the conformal block $\mathcal{F}^W$ and $f^W(x)$,
the negativity is constant as a function of $\ell/\beta$
for $x\sim 0$:
\begin{align}
  \label{s-channel neg}
  \mathscr{E}
  &=
    \displaystyle
    \frac{c}{2} \log\frac{\beta}{2\pi}, \quad x\to 0.
\end{align}
Note that the proper cut off factor is missing in this expression.
We simply replace $\beta/2\pi \to \beta/ (2\pi \epsilon C)$,
where as before $\epsilon$ is the UV cutoff, and $C$ is an unknown constant.
The negativity \eqref{s-channel neg}
can be matched with
the minimal entanglement wedge cross section
\eqref{entanglement wedge finite T}
by choosing $C$ properly ($C=2$).
On the other hand, with the solution $\mathcal{F}^+$
or $\mathcal{F}^-$, the negativity is not constant
for $x\sim 0$.

It is also worth while to have a look
at the series expansion \eqref{exp}:
with $h^{(2)}_n = a c$,
$h_n= \delta c$, 
the series expansion gives
\begin{align}
  \mathcal{F}(x)
  &=
  x^{- ac}
  \left[
  1 + \frac{a^2 c x}{2 \delta}
  +
  \cdots 
  \right].
\end{align}
In the replica limit
$a\to -c/8$ and $\delta =0$,
the each term in the expansion diverges,
except for the leading term.
Keeping this term alone,
and discarding (heuristically)
all divergent terms
reproduces $\mathcal{F}^W$.
Once again, this is a very similar situation as
the case of the two disjoint intervals.

\subsection{Thermofield Double State}

In this section, we consider the thermofield double state in CFT.
It is a purification of the mixed thermal state at
inverse temperature $\beta$ and given by
\begin{equation}
  \ket{{\it TFD}}
  =\frac{1}{\sqrt{Z(\beta)}} \sum_i e^{-\beta E_i /2} \ket{i}_1\ket{i}_2,
\end{equation}
where we have introduced the
two copies of the original CFT Hilbert space,
$\mathcal{H}_{{\it tot}} = \mathcal{H}_1 \otimes \mathcal{H}_2$,
and $|i\rangle_{1,2}$ is the $i$-th energy eigenstate with
energy $E_i$;
$Z(\beta)$ is the partition function.
When tracing out either copy of the CFT, the resulting reduced density matrix is thermal. The thermofield double state is conjectured to be dual to the AdS eternal black hole \cite{2003JHEP...04..021M}.

We follow Ref.~\cite{2014JHEP...10..060R} to obtain the negativity between the copies.
From the density matrix
\begin{equation}
    \rho^{\ }_{{\it TFD}} = \frac{1}{Z(\beta)} \sum_{i, j} e^{-\beta (E_i +E_j)/2} \ket{i}_1\bra{j}_1 \otimes \ket{i}_2\bra{j}_2,
\end{equation}
it is straightforward to compute
\begin{equation}
    \left| \rho_{{\it TFD}}^{T_1}\right|_1 = \frac{Z(\beta/2)^2}{Z(\beta)}.
\end{equation}
By taking the logarithm,
the entanglement negativity is given in terms of
the free energy
$F(\beta) = -(1/\beta)\log(Z(\beta))$ 
as
\begin{equation}
    \mathscr{E} (\rho^{\ }_{{\it TFD}}) = \beta [F(\beta) - F(\beta /2)].
\end{equation}

In the holographic pentagon code,
we can create the AdS eternal black hole
by connecting two codes with black holes at their center
by linking the black hole microstate legs (Fig.~\ref{happy_bh}).
Using \eqref{main} and
following our discussion in Sec.~\ref{Entanglement negativity in holographic perfect tensor network codes},
we see that 
the negativity is given by the 
area of the horizon $A_{{\it BH}}$ as
\begin{equation}
\mathscr{E}(\rho_{{\it TFD}}) \propto A_{{\it BH}},
\end{equation}
leading us to an interesting relation between the black hole area/entropy and the temperature/free energy of the dual CFT
\begin{equation}
\beta [F(\beta) - F(\beta /2)]_{{\it CFT}}\propto A_{BH}.
\end{equation}

When moving beyond the tensor network description to the full AdS/CFT,
we analogously find the minimal cross sectional area of the entanglement wedge to be the area of the black hole horizon. 
In ${\it AdS}_3/{\it CFT}_2$, 
we adopt the same normalization constant $(=3/2)$ as before relating
the negativity and the minimal entanglement wedge cross section.
This leads to
\begin{align}
  &
     \beta [F(\beta) - F(\beta/2)] = \frac{3}{2}\frac{A_{BH}}{4G_N}
    \\ \nonumber
 \mbox{or}\quad &    F(\beta) = F(\beta/2) + \frac{3A_{BH}}{8\beta G_N}.
\end{align}
We implement this recursively to obtain
\begin{equation}
    F(\beta) = \frac{3}{8G_N}\sum_{i=0}^{\infty}\frac{A_{BH}(\beta/2^i)}{\beta/2^i}.
\end{equation}
We work with the boundary of the Euclidean BTZ black hole
which is of length $L= 2\pi l_{{\it AdS}}$ (where $l_{{\it AdS}}$ is the radius
of AdS). 
Using $r_{{\it BH}} = {2\pi}/{\beta}$, we arrive at the formally divergent sum
\begin{equation}
    F(\beta) =  \frac{3\pi^2 l_{AdS}}{2 G_N \beta^2}\sum_{i = 0}^{\infty} 4^i.
\end{equation}
We can obtain a value for this by analytically continuing the geometric series. This gives us a value of $-1/3$. We use the Brown-Henneaux formula to arrive at 
\begin{equation}
  F(\beta) = \mathscr{O}
  \left(\frac{L}{\epsilon}\right) -\frac{\pi c L}{6 \beta^2} + \mathscr{O}(\epsilon)+\cdots,
\end{equation}
where $L$ is the size of the CFT system and $\epsilon$ is the cutoff. The
finite, universal part of the free energy precisely matches that for a thermal
CFT.

\section{Discussion}
\label{Discussion}

We have discussed
negativity in quantum error-correcting codes and tensor network models of holography.
We have shown that the entanglement negativity in these models
is captured by the minimal cross sectional area of
the entanglement wedge.
We have also conjectured a generalization to $AdS/CFT$ using the backreacted geometry of cosmic branes and have checked our proposal for a variety of configurations in ${\it AdS}_3/{\it CFT}_2$.

We close with a couple of discussions below.

\begin{figure}
    \includegraphics[width = 7cm]{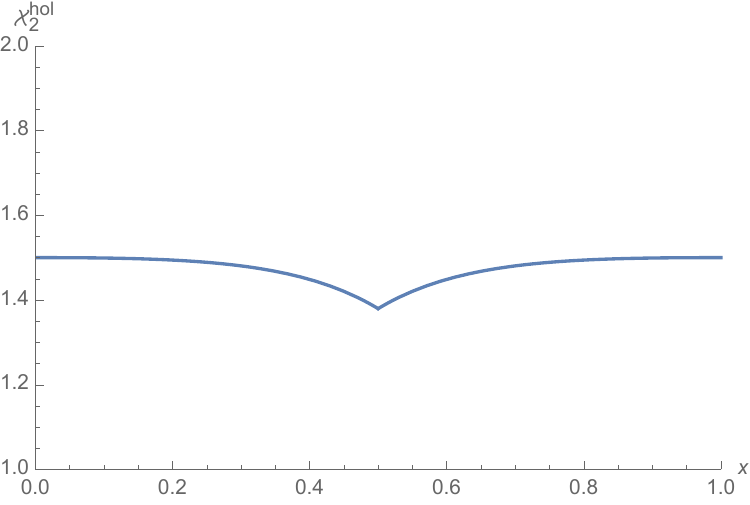}
    \caption{As we vary the cross-ratio, $x$,
      the proportionality between the negativity and area of the entanglement wedge cross section,
      $\mathcal{E}/S = S_{1/2}/S$,
      is perturbed from the value of $3/2$
      known for spherical entangling surfaces.
      Here, the R\'enyi entropy $S_{1/2}$ was computed using Zamolodchikov's recursion relation for Virasoro conformal blocks.}
    \label{backreaction}
\end{figure}

\paragraph{Non-spherical entangling surfaces}We stress that (\ref{sphere_neg})
should hold only for spherical entangling surfaces, which includes all examples
discussed in this paper so far. The backreaction in (\ref{general_formula})
becomes highly nontrivial when working with other geometries. For example, in
${\it AdS}_3/{\it CFT}_2$, if we bipartition the space into the union of two intervals and its complement, the entangling surface is no long a sphere (two points in this dimension).
Because we are working with the vacuum, we know that $\mathscr{E}_A = S_{1/2}(\rho_A)$.
As the cross-ratio is varied, the proportionality between the negativity and the area of the entanglement wedge cross-section changes (see Fig.
\ref{backreaction}).

\paragraph{Bit Threads}
We recall that the entanglement wedge is the bulk region corresponding to the reduced density matrix on the boundary. We can formulate the relation between the negativity and the entanglement wedge from the perspective of bit threads \cite{2017CMaPh.352..407F} by stating that the negativity between two boundary regions $A$ and $B$ is proportional to the maximum number of bit threads connecting the two regions through the bulk dual of $\rho_{AB}$. The maximization procedure is taken over all possible bit thread configurations. Unlike the case of entanglement entropy, the bit threads can no longer end on horizons. To account for non-spherically shaped entangling surfaces and R\'enyi entropies, it would be interesting to formulate bit threads in a language that could account for backreaction.

A similar picture can be made when considering entanglement of purification. This time the horizons represent the larger boundary Hilbert space needed to purify $\rho_{AB}$. In the effective bulk, there are no more horizons, so minimizing the maximum number of bit threads connecting the purified spaces of $A$ and $B$, is again proportional to the entanglement wedge cross section. If we are forced to use the horizons as the purifying Hilbert space, then the conjecture from Ref.~\cite{2018NatPh..14..573U} would be proven, though this is a highly nontrivial assumption.

Interestingly, explicit bit thread configurations in the entanglement wedge have been constructed in Ref.~\cite{2019JHEP...05..075A}. There, the bit threads were interpreted as the maximum number of Bell pairs that can be distilled from $\rho_{AB}$. This interpretation is extremely similar to that of logarithmic negativity which provides a bound on the distillable entanglement of mixed states \cite{2002PhRvA..65c2314V}. 

\paragraph{Covariant Conjecture} A natural covariant generalization may be considered in a similar way as the HRT formula. Here, we would need to find the proper analytic continuation of \textit{extremal} cosmic branes in the entanglement wedge. 

It would be fascinating to explore these generalizations quantitatively in order to better understand the connection between negativity and entanglement wedge cross sections.

\acknowledgments

We thank Chris Akers, Tom Faulkner, Ian MacCormack, Umang
Mehta, Masahiro Nozaki, Hassan Shapourian, Tadashi
Takayanagi, Mao Tian Tan, and Xueda Wen for useful
discussions. We thank Chris Akers for pointing out an error in an earlier version of this paper. SR is supported by a Simons Investigator
Grant from the Simons Foundation. 

\appendix

\section{Optional removal of horizon tensors}
\label{horizon}
We explain how to remove an additional layer of our tensor network and when this procedure is valid.
  We introduce the decompositions of the
  Hilbert spaces (both bulk and boundary)
  as follows.
  The boundary Hilbert space
  is decomposed into
  two parts,
  $\mathcal{H}_{AB}
  \otimes
  \mathcal{H}_C$.
  As for the bulk,
  there are degrees of freedom
  defined for the dangling points in
  the tensor network,
  as well as those living on bonds.
  The latter degrees of freedom correspond to
  $|\chi\rangle$ in the generic descriptions of
  subsystem QEC with complementary recovery.
  As for the ``dangling'' degrees of freedom,
  we decompose them as
  $\mathcal{H}_{b_{AB}}
  \otimes \mathcal{H}_{b_C}$
  where $b_{AB}$ represents the dangling Hilbert space 
  on the entanglement wedge of $AB$,
  whereas $b_C$ lives on the entanglement wedge of $C$.
  We further decompose $b_{AB}$ into
  $\tilde{b}_{AB}$ and $b_h$
  where 
  $b_h$ represents dangling degrees of freedom living
  on the ``horizon''; namely, we identify by the greedy
  algorithm, the minimal surface which cuts bonds
  connecting the entanglement wedge of $AB$ and $C$.
  $b_h$ are defined just inside of the horizon.
  We have a similar decomposition 
  of the Hilbert space associated to 
  the bulk link degrees of freedom.
  $\mathcal{H}_{l_{C}} \otimes \mathcal{H}_{l_{AB}} \otimes \mathcal{H}_{l_{e}}$
  where $l_{AB}$ represents the link Hilbert space 
  on the entanglement wedge of $AB$,
  whereas $l_C$ lives on the entanglement wedge of $C$,
  and finally, $l_{e}$ represents all links
  cut by the minimal surface.

  We are interested in the reduced density matrix $\rho_{AB}$
  on $\mathcal{H}_{AB}$
  (or $\tilde{\rho}_{AB}$ in the notation we used in QEC section).
  This is obtained from the total density matrix $\rho_{ABC}$
  on $\mathcal{H}_{ABC}$ by taking partial trace
  \begin{align}
    \rho_{AB} = \mathrm{Tr}_C\, \rho_{ABC}  
  \end{align}
  (For our situation, $\rho_{ABC}$ is pure.)

  By using the isometry $W$
  from $\mathcal{H}_{b_{C}}\otimes \mathcal{H}_{l_e}$ to
  $\mathcal{H}_C$,
  the reduced density matrix can be written as
  \begin{align}
    \rho_{AB} &= \mathrm{Tr}_C\, W \rho_{ABC} W^{\dag}
                \nonumber \\
     &= \mathrm{Tr}_{b_C, l_e}\, \rho_{AB, b_C l_e} 
  \end{align}
  where $\rho_{AB, b_C, l_e}$ is the result of the isometric map.
  The degrees of freedom $b_h$ are straightforward to trace over because $\rho_{AB, b_C, l_e}$ is a separable state
  \begin{align}
    \rho_{AB, b_C, l_e} =
  \sum_i p_i \rho^i_{AB, l_e}\otimes \rho^i_{b_C}
  \end{align}
  For example, if $\rho_{AB, b_C, l_e}$ is pure,
  \begin{align}
  \rho_{AB, b_C, l_e}
  =
  |\psi_{AB, b_C, l_e} \rangle\langle
  \psi_{AB,b_C, l_e} |
  \end{align}
  with 
  $|\psi_{AB,b_C, l_e}\rangle =
  |\psi_{AB, l_e}\rangle \otimes |\psi_{b_C}\rangle$,
  then, $\rho_{AB}$ is given by
  \begin{align}
    \label{step 1}
    \rho_{AB} =
    \mathrm{Tr}_{l_e}
  |\psi_{AB, l_e}\rangle 
    \langle  \psi_{AB, l_e}|.
  \end{align}

  For our purpose,
  we want to write $\rho_{AB}$
  using the degrees of freedom
  living on $b_h$. 
  We find this is possible
  under a certain condition, but not 
  in general.
  To state the condition,
  we focus on (for simplicity)
  the case where both the bulk
  state that we feed in to the QEC,
  and the boundary states are pure,
  and given by $|\psi\rangle_b$
  and $|\psi\rangle_{ABC}$, respectively.

  Recall that the tensor network (QEC) acts as
  an isometry from the (dangling) bulk to the boundary,
  i.e., there is an isometry relating
  $|\psi_b\rangle$ and $|\psi_{ABC}\rangle$.
  This means, in particular, if we Schmidt decompose
  $|\psi_b\rangle$ as
  \begin{align}
    |\psi_b\rangle
    =
    \sum_i c_i |\psi^i_{\tilde{b}_{AB},b_C} \rangle \otimes |\psi^i_{b_h}\rangle,
  \end{align}
  each term in the decomposition is mapped to a corresponding state
  $|\psi^i_{ABC}\rangle$, and hence we have a decomposition
  \begin{align}
    |\psi_{ABC} \rangle = \sum_i c_i |\psi^i_{ABC} \rangle.
  \end{align}

We engineer the state
\begin{align}
    |\phi_{ABC, b_h} \rangle = \sum_i c_i |\psi^i _{ABC}\rangle \otimes |\psi^i _{b_h}\rangle. 
\end{align}
  and assume it is a product state,
  \begin{align}
|\phi_{ABC, b_h} \rangle = |\psi _{ABC}\rangle \otimes |\psi _{b_h}\rangle. 
\label{condition}
  \end{align}
  This is our condition for removing the ``horizon layer" of the tensor network. Then, in this case, $\rho_{ABC}$ can be represented as a partial trace 
  over $b_h$:
  \begin{align}
    \rho_{ABC} = \mathrm{Tr}_{b_h}\, |\phi \rangle \langle \phi|_{ABC, b_h},
  \end{align}
  where $|\phi_{ABC, b_h}\rangle$
  is given by
  \begin{align}
    |\phi_{ABC, b_h} \rangle = |\psi \rangle_{ABC} \otimes |\psi \rangle_{b_h}. 
    \end{align}

    Now, for the case of this ``special class'' of
    bulk states, the tensor removal procedure by isometry
    can be repeated, 
    to reach \eqref{step 1},
    but since
     $\rho_{ABC}$ can now be written with a partial trace over $b_h$ of the engineered state
     $|\phi\rangle_{ABC,b_h}$, 
     \begin{align}
      \rho_{AB} =\mathrm{Tr}_{b_h} \mathrm{Tr}_{l_e}\,
      |\phi\rangle
      \langle  \phi|_{AB, l_e,b_h} 
      \end{align}
      where 
      $| \phi_{AB, l_e,b_h} \rangle$
      is obtained from
      $|\phi_{ABC, b_h}\rangle$
      by applying the isometry $W$.

      Applying an additional isometry,
      we can now remove degrees of freedom
      in $b_h$ and $l_e$. 
      After taking these partial traces,
      we are now left with 
      the description of 
      $\rho_{AB}$
      as the effective bulk state
      fed in to the (remaining)
      effective tensor network.
      In particular,
      the part of the effective tensor network
      that previously connected $\tilde{b}_{AB}$ and $b_h$
      can now be regarded as a horizon
      in the sense that we described before around Fig.~\ref{holographic_code};
      after removing $b_h$, these links are now dangling.

%

\end{document}